\documentstyle[12pt]{article}   

\large
\newcommand{\beq}{\begin{equation}}
\newcommand{\eeq}{\end{equation}}

\makeatletter
\@addtoreset{equation}{section}
\makeatother

\newtheorem{Theorem}{Theorem}[section]

\def\un{\underline}

\def\be{\begin{equation}}
\def\ee{\end{equation}}
\def\ba{\begin{eqnarray}}
\def\ea{\end{eqnarray}}

\def\A{{\cal A}}

\def\agb{{\overline {{\cal A}/{\cal G}}}}

\def\Ab{{\overline \A}}

\def\Comp{{\mathchoice
{\setbox0=\hbox{$\displaystyle\rm C$}\hbox{\hbox to0pt
{\kern0.4\wd0\vrule height0.9\ht0\hss}\box0}}
{\setbox0=\hbox{$\textstyle\rm C$}\hbox{\hbox to0pt
{\kern0.4\wd0\vrule height0.9\ht0\hss}\box0}}
{\setbox0=\hbox{$\scriptstyle\rm C$}\hbox{\hbox to0pt
{\kern0.4\wd0\vrule height0.9\ht0\hss}\box0}}
{\setbox0=\hbox{$\scriptscriptstyle\rm C$}\hbox{\hbox to0pt
{\kern0.4\wd0\vrule height0.9\ht0\hss}\box0}}}}
\def\Co{{\mathchoice
{\setbox0=\hbox{$\displaystyle\rm C$}\hbox{\hbox to0pt
{\kern0.4\wd0\vrule height0.9\ht0\hss}\box0}}
{\setbox0=\hbox{$\textstyle\rm C$}\hbox{\hbox to0pt
{\kern0.4\wd0\vrule height0.9\ht0\hss}\box0}}
{\setbox0=\hbox{$\scriptstyle\rm C$}\hbox{\hbox to0pt
{\kern0.4\wd0\vrule height0.9\ht0\hss}\box0}}
{\setbox0=\hbox{$\scriptscriptstyle\rm C$}\hbox{\hbox to0pt
{\kern0.4\wd0\vrule height0.9\ht0\hss}\box0}}}}
\def\Rl{{\mathchoice
{\setbox0=\hbox{$\displaystyle\rm R$}\hbox{\hbox to0pt
{\kern0.4\wd0\vrule height0.9\ht0\hss}\box0}}
{\setbox0=\hbox{$\textstyle\rm R$}\hbox{\hbox to0pt
{\kern0.4\wd0\vrule height0.9\ht0\hss}\box0}}
{\setbox0=\hbox{$\scriptstyle\rm R$}\hbox{\hbox to0pt
{\kern0.4\wd0\vrule height0.9\ht0\hss}\box0}}
{\setbox0=\hbox{$\scriptscriptstyle\rm R$}\hbox{\hbox to0pt
{\kern0.4\wd0\vrule height0.9\ht0\hss}\box0}}}}

\def\n{\nu}

\def\un{\underline}

\title{QSD V :\\ 
Quantum Gravity as the Natural Regulator of 
Matter Quantum Field Theories} 
\author{T. Thiemann\thanks{thiemann@math.harvard.edu}
\thanks{New Address : Albert-Einstein-Institut,
Max-Planck-Institut f\"ur Gravitationsphysik, Schlaatzweg 1, 14473 
Potsdam, Germany, Internet : thiemann@aei-potsdam.mpg.de}\\
       Physics Department, Harvard University, \\
       Cambridge, MA 02138, USA}
\date{{\small \today \\
       Preprint HUTMP-96/B-357}}

\begin{document}

\maketitle

\begin{abstract}
It is an old speculation in physics that, once the gravitational field is
successfully quantized, it should serve as the natural regulator of 
infrared and ultraviolet singularities that plague quantum field theories 
in a background metric.

We demonstrate that this idea is implemented in a precise sense within 
the framework of four-dimensional canonical Lorentzian quantum gravity in 
the continuum.

Specifically, we show that the Hamiltonian of the standard model supports
a representation in which finite linear combinations of Wilson loop
functionals around closed loops, as well as along open lines with fermionic
and Higgs field insertions at the end points are densely defined operators.

This Hamiltonian, surprisingly, does not suffer from any singularities, 
it is {\em completely finite} without renormalization. This property is 
shared by string theory. In contrast to string theory, however, we are 
dealing with a particular phase of the standard model coupled to gravity 
which is {\em entirely non-perturbatively} defined and {\em second 
quantized}. 
\end{abstract}

\section{Introduction}

It is an old idea in field theory that once the gravitational field has 
been successfully quantized then it should serve as the natural regulator
of matter quantum field theories. The argument is roughly that, since there 
is a fundamental length scale, namely the Planck scale
$\ell_p=\sqrt{\hbar\kappa}$ where $\kappa$ is Newton's constant, the 
gravitational field should serve
as an ultra-violet cut-off. The intuition coming from classical general 
relativity is that an elementary excitation
of the fields whose energy exceeds the Planck mass will have an energy
density that is high enough in order for the excitation to become a black
hole. Such tiny black holes, however, should evaporate within a Planck time
scale into excitations of lower energy. The conclusion is that elementary
excitations will have a 4-volume which is larger than or equal to
$\ell_p^4$.

In the present article we show a precise realization of this idea within
the framework of canonical quantization of matter-coupled four-dimensional
Lorentzian quantum gravity in the continuum. In brief, it is actually
possible to find a representation in which all matter Hamiltonians become,
after suitable regularization, densely defined operators. Upon removing
the regulator no renormalization is necessary and so the theory is, just like
string theory, {\em completely finite}. In contrast to string theory, 
however, our approach is fully non-perturbative and starts from the second
quantized field theory. In particular, our framework is mathematically 
completely rigorous, we have a well-defined Hilbert space and all the 
matter Hamiltonian operators are densely defined on it. Approximation 
schemes, if necessary to solve the theory, would not be formal perturbation
series with little if not no control on the error, but approximation
schemes with full control of convergence issues just like in usual 
quantum mechanics.

The intuitive picture that arises from the Hilbert space we choose is as 
follows :\\
The elementary excitations of the gravitational and gauge fields are 
concentrated along
open or closed strings while those of the fermion and scalar fields are 
located in the endpoints of the open strings. The Hamiltonian of the 
standard model and gravity, in this {\em diffeomorphism invariant phase} of 
the full theory, act by creating and annihilating those 
excitations which reminds of a {\em non-linear Fock representation}.

It should be stressed from the outset, however, that the 
string enters here as a completely kinematical object and unlike in string 
theory does not acquire any dynamical properties. It is just a label for 
the state, in fact, the same label that one uses for the familiar Wilson 
loop functionals that one knows from lattice gauge theory. 
Moreover, in contrast to string theory, the strings that acquire physical 
importance have necessarily singularities, i.e. they intersect in an 
arbitrarily complicated, non-differentiable, manner. \\ 
\\
The plan of the paper is as follows :

In section 2 we recall the quantum kinematics of the canonical 
approach  
from \cite{0,1,2,3,3a,4} for the gravitational and gauge sector and from
\cite{TTKin} for the Fermion and Higgs sector (see also \cite{RM-T,21}
for earlier work on the Fermion sector which, however, is described by a 
Hilbert space with an inappropriate inner product).

In section 3 we come to the regularization of the matter Hamiltonians which
is very similar in nature to the one performed for the gravitational field,
in fact the techniques used extend those introduced in \cite{5,6}. Roughly,
what we do is to introduce an ultraviolet cut-off by triangulating the 
spacelike hypersurfaces $\Sigma$ of the four-dimensional spacetime 
$M=\Rl\times\Sigma$ and then to 
take the continuum limit. We show that it exists and are able to precisely
display the action of the continnum operator. At no stage we encounter 
any singularities, these final operators do not require any renormalization.

In the first subsection we regulate the QCD Hamiltonian for any compact 
gauge group $G$.
Not surprisingly, the electric part of the Yang-Mills Hamiltonian becomes 
a sum of Laplace-Beltrami operators on $G$ and therefore is sensitive to 
the {\em colour} of the state while the magnetic part creates and 
annihilates new excitations, that is, it creates new Wilson loop functions.

In the next subsection we address the fermionic term. The fermionic 
Hamiltonian operator removes fermionic excitations from open string endpoints
and creates new excitations on the open string. Quite surprisingly,
these fermionic excitations are very different from those discussed in
\cite{RM-T,21} the underlying reason being the faithful implementation
of the reality conditions.

In the following subsection we discuss the regularization of the Higgs
Field Hamiltonian. The action of that Hamiltonian is analogous to the 
one of the gauge field. 

Finally, in the last subsection we outline a general procedure for 
regulating a rather general class of Hamiltonians whose corresponding 
Hamiltonian density has a density weight of one.

At this point the reader will be puzzled what role the gravitational 
field still plays. As will become obvious from the details, it enters the 
stage simultanously in two different ways (remember that the gravitational 
field couples 
to matter always through the three-metric $q_{ab}$ or the co-triad $e_a^i$
of $\Sigma$ (in addition, fermions couple to the extrinsic curvature as 
well)) : 
\begin{itemize} 
\item UV Singularities\\
Recall that one may control the usual ultraviolet singularities 
in terms of point-splitting regularizations of operator-valued 
distributions multiplied at the same point. For instance, we may have 
a singular square of operator-valued distributions of the form 
$\hat{F}(x)^2$ which one may regulate by point splitting $\int d^3y 
f_\epsilon(x-y)/\epsilon^3
\hat{F}(x)\hat{F}(y)$ and $\lim_{\epsilon\to 0}f_\epsilon(x)/\epsilon^3=
\delta(x)$. Notice that we automatically have broken diffeomorphism 
covariance because the points $x,y$ are split by a background metric. It 
turns out that 
the point splitting volume $\epsilon^3$ is {\em absorbed} by a certain 
gravitational operator, built from $q_{ab}$ that measures the volume of 
spatial regions. This 
is intuitively reasonable because the three-dimensional coordinate volume
$\epsilon^3$ cannot be measured by
a fixed background metric in a diffeomorphism invariant theory like general
relativity but must be measured by the dynamical metric $q_{ab}$ itself !
This volume operator therefore must enter the final expression of all matter
Hamiltonians. The formalism itself predicts how it enters, we do not have
to postulate this, of course, up to ordering ambiguities. Since this 
volume operator turns out to be densely defined on the Hilbert space
{\em the UV singularity} $1/\epsilon^3$ is removed by coupling 
quantum gravity, without renormalization, thereby reinstalling 
diffeomorphism covariance. 
\item IR Singularities :\\
This volume operator turns out to have a quite local action,
it vanishes everywhere except at points where the string ends or starts.
This is also an unexpected prediction of the formalism.
It is this feature which makes the Hamiltonian operators densely defined 
without that we have to introduce an infra-red (infinite volume) cut-off.\\
\end{itemize}
{\em In a sense, it is the volume operator which is the natural regulator 
of the matter quantum field theories by serving as a dynamical 
ultra-violet and infra-red cut-off !}\\ \\

And in accordance with what we said at the beginning of this section,
the volume is quantized with discrete spectrum, the quantum of volume 
being indeed of order $\ell_p^3$ (see \cite{12,13}) !
In particular, it will become obvious in the course of the construction 
that matter field theories are ultraviolet and infrared \\
a) {\em divergent without gravity} and need to be renormalized but are\\
b) {\em convergent with gravity} without that renormalization is necessary.\\
\\
In section 4 we perform various consistency checks on the theory, for 
instance, 
that we do not encounter quantum anomalies when computing commutators. 
This can be done only by restricting to the diffeomorphism 
invariant subspace of the Hilbert space in which we are interested only.
Expectedly, it is also in this context only that we can remove yet another
ultra-violet regulator for the connection field which enters in terms of a 
triangulation of $\Sigma$. When refining the triangulation ad infinitum 
we find that the continuum limit exists and yields well-defined operators on this
diffeomorphism invariant Hilbert space.

We also address the question whether the operators obtained are positive 
semi-definite, at least on the kinematical Hilbert space which, in view of 
some kind of ``quantum dominant energy 
condition", would be a re-assuring result because the matter Hamiltonian 
constraint plays the role of the timelike-timelike component of the 
energy momentum tensor. We find that, for each matter species separately, 
the answer is regularization and factor-ordering dependent.
We clarify the meaning of this result and point out that what only is
important is that the total (ADM) energy is non-negative (see \cite{19} 
for the pure gravity case).
Finally, we comment on the general construction of solutions to the 
full Hamiltonian constraint and demonstrate non-triviality of the theory
by displaying an uncountably infinite number of rigorous simple solutions.

In appendix A we derive the Dirac-Einstein canonical action in manifestly 
real form and in terms of the real connection variables that have proved 
successful in 
quantizing the source-free gravitational field in \cite{6}. This, to the 
best of our knowledge, has not yet been done in the literature.

\section{Preliminaries}

We begin by describing the field content of the theory.

The topology of the four-dimensional manifold is chosen, as always
in the canonical approach, to be $M=\Rl\times\Sigma$ where $\Sigma$
is a smooth 3-manifold which admits smooth Riemannian
metrics.  

On $\Sigma$, there is defined a co-triad field $e_a^i$ where $a,b,c,..$
denote tensorial indices and $i,j,k,..$ denote $su(2)$ indices. From
this field the 3-metric is derived via $q_{ab}=e_a^i e_b^i$. 
Furthermore, we have a field $K_a^i$ from which 
the extrinsic curvature $K_{ab}$ of $\Sigma$ is derived via
$K_{ab}=\mbox{sgn}(\det((e_c^j)))K_a^i e_b^i$. It turns out that the pair
$(K_a^i,E^a_i/\kappa)$ is a canonical one on the gravitational phase space 
where $E^a_i:=\frac{1}{2}\epsilon^{abc}\epsilon_{ijk}e_b^j e_c^k$. \\
The Hamiltonian constraint (or Wheeler-DeWitt constraint) of general 
relativity takes a quite complicated form 
in terms of these variables, thus it was natural to assume that for purposes
of quantizing canonical gravity it is mandatory to cast the theory into
polynomial form. The famous discovery due to Ashtekar \cite{7} is that
this indeed possible by performing a certain canonical transformation
on the gravitational phase space. However, this transformation comes at
two prizes :\\ 
1) The Hamiltonian constraint is polynomial only after rescaling it by
$\sqrt{\det((q_{ab}))}$. This is bad because the constraint now adopts
a density weight of two which rules out a diffeomorphism covariant
regularization and will require a troubelsome multiplicative 
renormalization. This will become apparent in section 3.\\
2) The so-called canonical Ashtekar variables $(^\Co A_a^i=\Gamma_a^i-
iK_a^i,iE^a_i/\kappa)$, where $\Gamma_a^i$ is the spin-connection of
$e_a^i$, are complex-valued. This is bad because 
the Ashtekar connection $^\Co A_a^i$ 
is the connection of a principal $SL(2,\Co)$ bundle, that is, the gauge
group is non-compact and makes the rich arsenal of techniques that 
have been developed for gauge theories with compact gauge group
inaccessible.\\
There have been two quite different proposals to deal with problem 2).
First of all, in \cite{8} it was pointed out that one can also use
a {\em real-valued} Ashtekar connection $A_a^i=\Gamma_a^i+K_a^i$ at the 
prize of living with a fairly complicated Lorentzian Hamiltonian constraint.
The virtue is that this at least opens access to the techniques developed in
\cite{0,1,2,3,3a,4} and equips us with a Hilbert space structure that 
faithfully
implements the reality conditions. Restricted to Euclidean gravity this was
also proposed in \cite{9}.\\
The second proposal is to perform a Wick rotation on the canonical phase
space \cite{10}. The virtue of this is that one can start by quantizing
the Euclidean Hamiltonian constraint in terms of the real canonical variables
$(A_a^i,E^a_i/\kappa)$ which takes care of the reality structure of the
theory and keeps the constraint polynomial. In a second step then one
would perform a Bargman-Segal kind of transform to the Lorentzian theory
described by complex valued connections (compare also a modified
procedure \cite{11} which could enable one to stay purely within a real
connection theory). The drawback is that the 
generator of the Wick transform adopts a quite complicated form which made 
it hard to imagine how one would be able to quantize it (see, however,
\cite{6} for a proposal for a self-adjoint operator).\\
Apart from the problems mentioned, both proposals still suffer from
the problem 1) described above.\\   
In \cite{5,6} a novel technique was introduced which solves both problems
1),2) in one stroke and on top defines the generator of the Wick rotation
transform. The resulting Lorentzian Hamiltonian constraint is densely
defined, anomaly-free and one has a Hilbert space that incorporates the
correct reality conditions.\\
This paper is devoted to the extension of this technique to the 
non-gravitational sector.\\

Let $G$ be an arbitrary compact gauge group, for instance
the gauge group of the standard model. Denote by $I,J,K,...$ $Lie(G)$
indices. We introduce classical Grassman-valued spinor fields 
$\eta=(\eta_{A,\mu})$
where $A,B,C,..$ denote indices associated with the gravitational $SU(2)$
and $\mu,\nu,\rho,..$ with the group $G$. The fermion species 
$\eta$ transforms like a
scalar and according to an irreducible representation of $SU(2)\times G$.
It turns out that in its manifestly real form (the associated conjugation
is just complex conjugation for non-spinorial variables and for spinorial
fields it involves a cyclic reversal of order in products) the most 
convenient description of the constraints is in terms of half-densities
$\xi:=\sqrt[4]{\det(q)} \eta$. The momentum conjugate to $\xi_{A,\mu}$ is
then just given by $\pi_{A,\mu}=i\overline{\xi}_{A,\mu}$ and the 
real-valued gravitational connection is given by $A_a^i=\Gamma_a^i+K_a^i$
just as in the source free case.
As we will see in appendix A, the connection is real only if we use the 
quantities $\xi$ with density weight $1/2$, if we would use the scalar
variables $\eta$ as in \cite{21} then the gravitational connection 
becomes by the argument given in \cite{Jacobson} 
\be \label{0}
 A_a^i=\Gamma_a^i+K_a^i+\frac{i}{4\sqrt{\det(q)}}e_a^i
\overline{\xi}_{A,\mu}\xi_{A,\mu} 
\ee
which is complex valued and therefore makes the techniques in 
\cite{1,2,3,3a,4} inaccessable.\\
Notice that it is no lack of generality to restrict ourselves to just one
kind of helicity : If we have several fermion species then we can always 
perform the canonical transformation 
$(i\bar{\xi},\xi)\to(i\epsilon\xi,\overline{\epsilon\xi})$ where 
$\epsilon$ is the spinor-metric, the totally skew symbol in two 
dimensions. Notice that there is no 
minus sign missing because we take the fermion fields to be anti-commuting,
the action is form-invariant under this transformation \cite{Jacobson}.

In the gauge sector we have canonical pairs 
$(\underline{A}_a^I,\underline{E}^a_I/Q^2)$ where the first entry is a $G$
connection and the second entry is the associated electric field, $Q$ is the
Yang-Mills coupling constant.
Finally, we may have scalar Higgs fields described by a canonical pair 
$(\phi_I,p^I)$ transforming according to the adjoint representation of $G$. 
Without loss of generality we can take these as real valued by suitably 
raising the number of Higgs families. Here and 
in what follows we assume that indices $I,J,K,..$ are raised and lowered 
with the Cartan-Killing metric $\delta_{IJ}$ of $G$ which we take to be 
semi-simple up to factors of $U(1)$.

We could also introduce Rarity-Schwinger fields and make everything
supersymmetric but since this will not add new features as compared to the
ordinary spinorial action, we refrain from doing so.\\
With this field content we then have the 
following Lorentzian Hamiltonian constraints :
\ba \label{1}
H_{Einstein}&=&\frac{1}{\kappa\sqrt{\det(q)}}
\mbox{tr}(2\{[K_a,K_b]-F_{ab}\}[E^a,E^b])+\lambda\sqrt{\det(q)}
\nonumber\\
H_{Dirac}&=&E^a_i\frac{1}{2\sqrt{\det(q)}}[i\pi^T\tau_i{\cal D}_a\xi+
{\cal D}_a(\pi^T\tau_i\xi)+\frac{i}{2}K_a^j\pi^T\xi+c.c.]
\nonumber\\
H_{YM}&=&\frac{q_{ab}}{2 Q^2\sqrt{\det(q)}}
[\underline{E}^a_I\underline{E}^b_I
+\underline{B}^a_I\underline{B}^b_I]
\nonumber\\
H_{Higgs}&=&\frac{1}{2}
(\frac{p^I p^I}{\kappa\sqrt{\det(q)}}+\sqrt{\det(q)}
[q^{ab}({\cal D}_a\phi_I)({\cal
D}_b\phi_I)/\kappa+P(\phi_I\phi_I)/(\hbar\kappa^2)]).
\ea
Here we have denoted by $\tau_i$ the generators of the Lie algebra of 
$su(2)$ with the convention $[\tau_i,\tau_j]=\epsilon_{ijk}\tau_k$,
$F_{ab}$ is the curvature of $A_a$ (one can check that all the constraints
remain form-invariant under the canonical transformation that turns 
the fermions into half-densities, see appendix A), $\cal D$ is the 
covariant 
derivative with respect to $SU(2)\times G$, that is, with respect to 
$\omega_a:=A_a+\underline{A}_a$
and $\underline{B}^a$ is the magnetic field of the Yang-Mills connection. 
We have included a cosmological constant ($\lambda$) and $P$ denotes an 
arbitrarily chosen gauge invariant function of the Higgs field (not 
including spatial
derivatives), the Higgs potential. Notice that we have rescaled the
Higgs field by $\sqrt{\kappa}$ in order to make it dimensionless.\\
The unfamiliar terms in the Dirac Hamiltonian proportional to the total 
derivative and $K_a^i$ arise because 1) we are dealing with half densities 
rather than scalars and 2) we couple the real connection $A$ to the 
spinor fields while in the traditional approach it is naturally the complex
valued (anti-)self-dual part of the spatial projection of the 
spin-connection that couples to them. Thus, these additional terms are 
the required correction terms if we describe the theory in the variables 
we chose. The interested reader is referred to appendix A in order to see
how these corrections come about. As usual, the ``c.c." means involution
(complex conjugation for complex valued fields and an additional reversal 
of order is implied for the Grassman valued fields).\\
In (\ref{1}) we have written only one family member of the possibly arbitrary
large family of field species, in particular, we can have an arbitrary number
of gauge fields all associated with different gauge groups and associated 
``quarks" and ``Higgs" fields and transforming under different 
irreducible representations of $SU(2)\times G$. However, we will not deal 
with these straightforward generalizations and consider only one species of
fermions or Higgs fields respectively which transform under the fundamental
representation of both $SU(2)$ and $G$ or the adjoint representation of $G$ 
respectively. Also, one could easily deal with a 
more complicated ``unified gauge group" which is not of the product type
$SU(2)\times G$ but contains it as a subgroup. However, for simplicity 
and because one does not expect a unification of the gauge group of the 
standard model and the gauge group underlying the frame bundle, we 
refrain also from treating this more general case. \\
This furnishes the description of the classical field content.\\
\\
We now come to the quantum theory. We can immediately apply the techniques
of \cite{0,1,2,3,3a,4} to write down a kinematical inner product for the
gravitational and Yang-Mills sector that faithfully incorporates all the 
reality conditions. We get a Hilbert space 
$L_2(\Ab_{SU(2)}\times\Ab_G,d\mu_{AL,SU(2)}\otimes d\mu_{AL,G})$ where the 
index ``AL" stands for Ashtekar-Lewandowski measure and the group index
indicates to which gauge group the Ashtekar-Lewandowski measure is 
assigned. The reader interested
in the constructions and techniques around the space of generalized
connections modulo gauge transformations is urged to consult the papers
listed. In particular, the probability measure $\mu_{AL}$ is very natural and 
diffeomorphism invariant. If we are interested in gauge invariant 
functions of connections alone, then the space of generalized connections
$\Ab$ can be replaced by the space of generalized connections modulo gauge 
transformations $\agb$.

The extension of the framework to Higgs and fermionic fields is not 
entirely straightforward :\\
Let us first focus on the Higgs field. 
Assume that we choose $\phi_I(x)$ as our basic configuration field variable.
As argued in \cite{TTKin}, in a diffeomorphism invariant theory this 
assumption has consequences which leads to inconsistencies. Basically, the 
problem is the following : The variables $\phi_I(x)$ are real-valued 
and thus there does not exist a translation invariant measure on the space
of these $\phi_I$'s. For a quantum field theory in a fixed background there
is no problem, a natural kinematical measure that incorporates the 
reality conditions is a Gaussian measure leading to a usual Fock Hilbert 
space. However, a Gaussian measure for a scalar field,
rigorously defined through its covariance, is {\em always background 
dependent} or, in other words, cannot be diffeomorphism invariant ! An 
intuitive way to see this is by recalling that the covariance is 
determined by the characteristic functional $exp(\int d^3x\int d^3y 
C_{IJ}(x,y) f^I(x) f^J(y))$ of the 
measure which in turn is the expectation value of $\exp(i\int d^3x f^I(x) 
\phi_I(x))$ where $f^I$ are some test functions. However, the fact that
$\phi_I$ is a scalar implies that the kernel $C_{IJ}(x,y)$ of the covariance 
is a density of weight one and therefore the characteristic functional is  
background dependent. See \cite{TTKin} for more details.

Thus we need a new approach which does not use a 
Gaussian measure and therefore we must not use $\phi_I$ as a basic 
variable but some variable that is valued in a bounded set.
This motivates to use the variables $U(v):=\exp(\phi_I(v)\tau_I)$ quite 
in analogy with the holonomy for a connection and we will call them 
``point holonomies". Point holonomies are $G$-valued and, since $G$ is 
compact, its matrix elements are therefore bounded. 
In \cite{TTKin} we 
construct a representation in which the $U(v)$ are promoted to unitary
operators (since we can replace $G$ by a unitary group by the theorem
due to Weyl that any compact group is equivalent to a unitary one).
If we are dealing not with a Higgs field but just with a real scalar
field then we may use $U(v)=e^{i\phi(v)}$. The Hilbert space to be used
is surprisingly simple to describe : there is a certain space 
$\overline{{\cal U}}$
of generalized Higgs fields which 
turns out to be in bijection with $\mbox{Fun}(\Sigma,G)$, the space of
{\em all} functions from $\Sigma$ to the gauge group. That is, a typical 
such function is a ``wild",
arbitrarily discontinuous function, it is a wild Higgs field. On that 
space we have a measure
$\mu_U$ which is a rigorously defined $\sigma$-additive probability 
measure on $\overline{{\cal U}}$ which is formally given by the uncountable 
direct product $d\mu_U(\phi):=\prod_{v\in\Sigma}d\mu_H(U(v))$ where
$\mu_H$ denotes the Haar measure on $G$. The Hilbert space is then the 
corresponding $L_2(\overline{{\cal U}},d\mu_U)$ space and one can show
\cite{TTKin}
that this is the unique Hilbert space selected by the adjointness 
relations, once we have chosen the space $\overline{{\cal U}}$ as the 
quantum configuration space. Expectedly, the mathematical description is 
very similar to the one for gauge fields \cite{0,1,2,3,3a,4}.

Next we come to the fermion fields $\xi$ which, as explained above, have 
density weight $1/2$.\\
It turns out \cite{TTKin} that the faithful implementation of the reality 
conditions forces us to work in a representation in which the objects
$$
\theta_{A\mu}(x):=\int_\Sigma d^3y \sqrt{\delta(x,y)} \xi_{A\mu}(y)
:=\lim_{\epsilon\to 0} \int_\Sigma d^3y 
\frac{\chi_\epsilon(x,y)}{\sqrt{\epsilon^3}} \xi_{A\mu}(y)
$$
become densely defined multiplication operators. Here $\chi_\epsilon(x,y)$
is the characteristic function of a box of Lebesgue measure $\epsilon^3$ 
and center $x$. The $\theta$ are by inspection scalar Grassman-valued 
quantities because the $\delta$ distribution is a density of weight one.
In calculations it is understood that the 
$\epsilon\to 0$ limit is performed only after the manipulation under 
consideration is performed \cite{TTKin}.\\
Consider then the $n=2d$ Grassman variables
$\theta_i(v),\; A=1,2,\;\mu=1,..,d$ where $d$ denotes the dimension of 
the fundamental representation of $G$. Here we have have introduced a
compound symbol $i$ instead of $A\mu$ to simplify the notation. These 
variables coordinatize together with their conjugates the superspace $S_v$
at point $v$.
Since Grassman fields anti-commute, any product of more than $n$ of these
$\theta_i(v),\;i=1,..,n$ will vanish. The vector space of monomials 
of order $k$ is $n!/(k!(n-k)!)$ dimensional where $k=0,1,..,n$ and the 
full vector space $Q_v$ built from all monomials has dimension $2^n$.
The quantum configuration space is the uncountable direct
product (``superspace") $\overline{{\cal S}}:=\prod_{v\in\Sigma} S_v$ and in 
order to define an inner product on $\overline{{\cal S}}$ it turns out
to be sufficient to define an inner product on each $S_v$ coming from a 
probability ``measure".
The ``measure" on $S_v$ is a modified form of the Berezin symbolic integral
\cite{Berezin} :
$$ 
dm(\overline{\theta},\theta)=d\overline{\theta}d\theta 
e^{\overline{\theta}\theta} \mbox{ and }
dm_v=\otimes_{i=1}^n dm(\overline{\theta}_i(v),
\theta_i(v)).
$$
The fermionic Hilbert space is then simply given by 
$$ 
{\cal H}_F=L_2(\overline{{\cal S}},d\mu_F)=
\otimes_{v\in\Sigma} L_2(S_v,dm_v)
$$
where ``F" stands for fermionic and it is understood that we
integrate only linear combinarions of functions of the form $\bar{f}g$ where
$f,g$ are both {\em holomorphic} (that is, a function on $\bar{S}$ which 
depends on $\theta_i(v)$ only but not on $\bar{\theta}_i(v)$).  
As a result, the integral of any function of the type $f^\star f$, where 
$f$ is any holomorphic function, is strictly positive and so we have an 
inner product. This inner product, when restricted to one point $v$, is  
easily seen to be the standard inner product on $Q_v$ when viewed as the 
vector space of exteriour forms of maximal degree $n$.
Thus, ${\cal H}_F$ is a space of holomorphic square integrable 
functions on $\bar{S}$ with respect to $d\mu_F$.
The Fermion measure $\mu_F$ is easily seen 
to be gauge and diffeomorphism invariant. \\
The reader is referred to \cite{TTKin} for 
a more complete treatment where it is also shown that the reality condition
$\xi^\star=-i\pi$ is faithfully implemented in the inner product. 
The reader will find there also an extension of the 
framework to the diffeomorphism invariant subspace of the Hilbert space.\\
\\
Let us summarize : the Hilbert space of (not necessarily gauge invariant)
functions of gravitational, gauge, spinor and Higgs fields is given by
$${\cal H}:=
L_2(\Ab_{SU(2)},d\mu_{AL}(SU(2)))\otimes
L_2(\Ab_G,d\mu_{AL}(G))\otimes L_2(\overline{{\cal S}},d\mu_F)
\otimes L_2(\overline{{\cal U}},d\mu_U). $$
The Hilbert space of gauge invariant functions will be just the restriction
of $\cal H$ to gauge invariant functions. It turns out that, because 
our total measure is a probability measure, gauge invariant functions 
will be still integrable with respect to it, in other words, ``the gauge 
group volume" equals unity in our case !

A natural gauge invariant object associated with spinor fields, Higgs 
fields and gauge 
fields are ``spin-colour-network states" \cite{TTKin}. By this we mean the 
following :
Let $\gamma$ be a piecewise analytic graph with edges $e$ and vertices $v$
which is not necessarily connected or closed.
By suitably subdividing edges into two halves we can assume that all edges
are outgoing at a vertex.
Given a (generalized) connection $\omega_a=A_a+\underline{A}_a$
we can compute the holonomies 
$h_e(A),\;\underline{h}_e(\underline{A}),H_e(\omega)=h_e(A)\underline{h}_e(
\underline{A})$. With each edge $e$ we associate a spin $j_e$ and a
colour $c_e$ corresponding to irreducible representations of $SU(2)$ and 
of $G$ respectively (for instance for $G=SU(N)$, $c_e$ is an array
of $N-1$ not increasing integers corresponding to the frame of a 
Young diagramme). Furthermore, with each vertex $v\in V(\gamma)$ we 
associate an integer
$n_v$, yet another colour $C_v$ and two projectors $p_v,q_v$. 
Here $V(\gamma)$ denotes the set of vertices of $\gamma$.
The integer 
$n_v$ corresponds to the subvector space of $Q_v$ spanned by monomials of
degree $n_v$. \\
Likewise, the colour $C_v$
stands for an irreducible representation of $G$, evaluated at the point 
holonomy $U(v)$. 
The projector $p_v$ is a certain $SU(2)$ invariant matrix which projects 
onto one of the linearly independent 
trivial representations contained in the decomposition into irreducibles
of the tensor product consisting of\\
a) the $n_v-$fold tensor product 
of fundamental representations of $SU(2)$ associated with the 
subvector space of $Q_v$ spanned by the monomials of degree $n_v$ and\\
b) the tensor product of the
irreducible representations $j_e$ of $SU(2)$ of spin $j_e$ where $e$ runs 
through the subset of edges of $\gamma$ which start at $v$.\\ 
Likewise,
the projector $q_v$, repeats the same procedure just that $SU(2)$
is being replaced by $G$ and that we need to consider in addition the 
adjoint representation associated with $C_v$ coming from the Higgs field at
$v$. 
Now we simply contract all the indices of the tensor product of\\ 
1) the irreducible representations evaluated at the holonomy of the given 
connection, \\
2) the fundamental representations evaluated at the given spinor field 
and\\ 
3) the adjoint representations evaluated at the given scalar field,\\
all associated with the same vertex $v$, with the projectors $p_v,q_v$ in 
the obvious way and for all $v\in V(\gamma)$. The result is a 
gauge invariant state 
$$ T_{\gamma,[\vec{j},\vec{n},\vec{p}],[\vec{c},\vec{C},\vec{q}]}$$
which we will call a spin-colour-network states because they extend the 
definition of the pure spin-network states which arise in the source-free
case (e.g. \cite{4}).\\
These spin-colour-networks turn out to be a basis for the subspace of 
gauge invariant functions. They are not orthonormal, but almost : we just
need to decompose the fermionic dependence into an orthonormal 
basis for each of the $Q_v$ \cite{TTKin}.\\
\\
This furnishes the summary of the quantum kinematics. We now turn to the 
quantum dynamics.

\section{Regularization}

The regularization of the Wheeler-DeWitt Hamiltonian constraint was carried
out in \cite{5,6}. We therefore can focus on the remaining Hamiltonians.

\subsection{Gauge sector}

We begin by looking at the electric piece 
$Q^2 H^E_{YM}(N)=\int d^3x N\frac{q_{ab}}{2\sqrt{\det(q)}}
\underline{E}^a_I \underline{E}^b_I$.
Recall from \cite{5,6} that the following identity was key
\be \label{2}
\frac{1}{\kappa}\{A_a^i,V\}
=\frac{\delta V}{\delta E^a_i}=2\mbox{sgn}(\det((e_b^j)))e_a^i
\ee
where $V=\int d^3x \sqrt{\det(q)}$ is the total volume of the hypersurface
(in the asymptotically flat case the appropriate definition of the 
functional derivative of $V$ involves a certain limiting procedure
\cite{5}).\\
Let now $\epsilon$ be a small number and let $\chi_\epsilon(x,y)
=\prod_{a=1}^3 \theta(\epsilon/2-|x^a-y^a|)$ be the characteristic function
of a cube of coordinate volume $\epsilon^3$ with center $x$. That is, 
we have chosen some frame andd therefore broken diffeomorphism 
covariance in the regularization step. We are going to remove the 
regulator later again and also recover diffeomorphism covariance. Also let
$V(x,\epsilon):=\int d^3y \chi_\epsilon(x,y)\sqrt{\det(q)}$ be the 
volume of that box as measured by $q_{ab}$. Then, since
$\lim_{\epsilon\to 0}\frac{1}{\epsilon^3}\chi_\epsilon(x,y)=\delta(x,y)$
we have $\lim_{\epsilon\to0}
\frac{1}{\epsilon^3}V(x,\epsilon)=\sqrt{\det(q)}(x)$. 
It is also easy to see that for each $\epsilon>0$ we have that
$\delta V/\delta E^a_i(x)=\delta V(x,\epsilon)/\delta E^a_i(x)$.
The simple trick that we are going to use is as follows : let $f,g$ be 
some integrable functions on $\Sigma$ with respect to Lebesgue measure.
Thus
$f(x)=\lim_{\epsilon\to 0}\frac{1}{\epsilon^3}\int d^3y f(y)\chi_\epsilon(x,y)
=:\lim_{\epsilon\to 0}\frac{1}{\epsilon^3}f(x,\epsilon)$ and similar for 
$g$. Then 
$\lim_{\epsilon\to 0}[f(x,\epsilon)/g(y,\epsilon)]=\lim_{\epsilon\to 0} 
[\{f(x,\epsilon)/\epsilon^3\}/\{g(y,\epsilon)/\epsilon^3\}]=
f(x)/g(y)$, that is, the two singular factors of $\epsilon^3$ cancel each 
other in the quotient.\\
With this preparation it 
follows that we have the following classical identity 
\ba \label{3}
&& 2\kappa^2 Q^2 H^E_{YM}(N)\nonumber\\
&=&\lim_{\epsilon\to 0}\frac{1}{\epsilon^3}
\int d^3x N(x)\frac{\{A_a^i(x),V\}}{2\sqrt[4]{\det(q)}(x)} \un{E}^a_I(x)
\int d^3y \chi_\epsilon(x,y) \frac{\{A_b^i(y),V\}}{2\sqrt[4]{\det(q)}(y)}
\un{E}^b_I(y)
\nonumber\\
&=&\lim_{\epsilon\to 0}\frac{1}{\epsilon^3}
\int d^3x N(x)\frac{\{A_a^i(x),V(x,\epsilon)\}}{2\sqrt[4]{\det(q)}(x)} 
\un{E}^a_I(x) \int d^3y \chi_\epsilon(x,y) 
\frac{\{A_b^i(y),V(y,\epsilon)\}}{2\sqrt[4]{\det(q)}(y)}\un{E}^b_I(y)
\nonumber\\
&=&\lim_{\epsilon\to 0}\frac{1}{\epsilon^3}
\int d^3x N(x)
\frac{\{A_a^i(x),V(x,\epsilon)\}}{2\sqrt{\frac{1}{\epsilon^3}V(x,\epsilon)}}
\un{E}^a_I(x) \int d^3y \chi_\epsilon(x,y) 
\frac{\{A_b^i(y),V(y,\epsilon)\}}{2\sqrt{\frac{1}{\epsilon^3}V(y,\epsilon)}}
\un{E}^b_I(y)
\nonumber\\
&=&\lim_{\epsilon\to 0}
\int d^3x N(x)
\frac{\{A_a^i(x),V(x,\epsilon)\}}{2\sqrt{V(x,\epsilon)}}
\un{E}^a_I(x) \int d^3y \chi_\epsilon(x,y) 
\frac{\{A_b^i(y),V(y,\epsilon)\}}{2\sqrt{V(y,\epsilon)}}
\un{E}^b_I(y)
\nonumber\\
&=&\lim_{\epsilon\to 0} \int d^3x N(x)
\{A_a^i(x),\sqrt{V(x,\epsilon)}\} \un{E}^a_I(x) \int d^3y \chi_\epsilon(x,y) 
\{A_b^i(y),\sqrt{V(y,\epsilon)}\} \un{E}^b_I(y)\nonumber\\
&&
\ea
which demonstrates that we can neatly absorb the annoying $1/\sqrt{\det{q}}$
into a Poisson bracket, of course at the prize of breaking gauge invariance
at finite $\epsilon$. The removal of the divergent factor $1/\epsilon^3$
has occured precisely because we kept the density weight of the constraint
to be one !\\
We have similarily for the magnetic term 
\ba \label{4}
2 Q^2\kappa^2 H^B_{YM}(N)&=&\lim_{\epsilon\to 0}
\int d^3x N(x)
\{A_a^i(x),\sqrt{V(x,\epsilon)}\} \un{B}^a_I(x) \int d^3y \chi_\epsilon(x,y) 
\times\nonumber\\&\times&  
\{A_b^i(y),\sqrt{V(y,\epsilon)}\} \un{B}^b_I(y)\;.
\ea
We come now to the quantization of the Yang-Mills Hamiltonian constraint.
This will be somewhat different for the electric and the magnetic part
so that we describe them separately. Notice that we have no factor ordering
problem at all as far as the question, whether to order the gravitational or
the gauge theory variables to the left or to the right,
is concerned.\\
Let us then start with the electric part. We choose to order the Yang Mills
electric fields to the right and replace $\un{E}^a_I\to-i\hbar \delta/
\delta\un{A}_a^I,\;V\to\hat{V}$ ($\hat{V}(R)$, for an arbitrary region
$R$, was described in \cite{12,13,14,15}) and Poisson brackets
by commutators times $1/i\hbar$. If $\ell_p=\sqrt{\hbar\kappa},m_p=
\sqrt{\hbar/\kappa}$ denote
Planck length and mass respectively then we obtain on a function $f$ 
cylindrical with respect 
to a graph $\gamma$ the following result ($\alpha_Q=\hbar Q^2$ is the 
dimensionless fine structure constant)\footnote{here and in the 
regularizations that follow we are going to apply the operator first only
to functions of classical (i.e. smooth) fields and then extend the end 
result to the quantum configuration space} 
\ba \label{5} 
&&-\hat{H}^{E,\epsilon}_{YM}(N)f\nonumber\\
&=&\sum_{e,e'}\frac{\alpha_Q m_p}{2\ell_p^3}\int d^3x N(x)
[A_a^i(x),\sqrt{\hat{V}(x,\epsilon)}]\int d^3y \chi_\epsilon(x,y) 
[A_b^i(y),\sqrt{\hat{V}(y,\epsilon)}]\times\nonumber\\
&\times&\int_0^1 dt\delta(x,e(t)) \dot{e}^a(t)
\int_0^1 dt'\delta(y,e'(t')) \dot{e}^{\prime b}(t')\times\nonumber\\
&\times&\{[
\mbox{tr}(\un{h}_e(0,t)\tau_I\un{h}_e(t,1)\partial/\partial\un{h}_e(0,1))
\mbox{tr}(\un{h}_{e'}(0,t')\tau_I\un{h}_{e'}(t',1)
\partial/\partial\un{h}_{e'}(0,1))]
\nonumber\\
&&+\delta_{e,e'}[\theta(t'-t)
\mbox{tr}(\un{h}_e(0,t)\tau_I\un{h}_e(t,t')\tau_I\un{h}_e(t',1)
\partial/\partial\un{h}_e(0,1))
\nonumber\\
&& +\theta(t-t')
\mbox{tr}(\un{h}_e(0,t')\tau_I\un{h}_e(t',t)\tau_I\un{h}_e(t,1)
\partial/\partial\un{h}_e(0,1))]\}f
\nonumber\\
&=& \sum_{e,e'}\frac{\alpha_Q m_p}{2\ell_p^3}
\int_0^1 dt \int_0^1 dt' N(e(t))\chi_\epsilon(e(t),e'(t')) 
\times\nonumber\\&\times&  
[A_a^i(e(t))\dot{e}^a(t),\sqrt{\hat{V}((e(t),\epsilon)}]
[A_b^i(e'(t'))\dot{e}^{\prime b}(t'),\sqrt{\hat{V}(e'(t'),\epsilon)}]
\times\nonumber\\
&\times&\{[
\mbox{tr}(\un{h}_e(0,t)\tau_I\un{h}_e(t,1)\partial/\partial\un{h}_e(0,1))
\mbox{tr}(\un{h}_{e'}(0,t')\tau_I\un{h}_{e'}(t',1)
\partial/\partial\un{h}_{e'}(0,1))]
\nonumber\\
&&+\delta_{e,e'}[\theta(t'-t)
\mbox{tr}(\un{h}_e(0,t)\tau_I\un{h}_e(t,t')\tau_I\un{h}_e(t',1)
\partial/\partial\un{h}_e(0,1)) \nonumber\\ 
&&+\theta(t-t')
\mbox{tr}(\un{h}_e(0,t')\tau_I\un{h}_e(t',t)\tau_I\un{h}_e(t,1)
\partial/\partial\un{h}_e(0,1))]\}f.
\ea
Here we have used the step functions $\theta(t)=1$ if $t>0$ and $0$ 
otherwise. The negative sign in (\ref{5}) stems from the $(-i)^2$ coming 
from the two factors of the electrical field.\\
The next step consists in replacing the integrals by Riemann sums, suggested
by the expansion $[h_e(t,t+\delta t),\hat{O}]=\delta
t\dot{e}^a(t)[A_a(e(t)),\hat{O}]+o(\delta t^2)$ for an arbitrary operator
$\hat{O}$. We choose an arbitrary  partition of the interval $[0,1]$ into 
$n$ intervals with endpoints $t_k,\;k=0,..,n$ which we can since the 
Riemann integral is independent of the partition that defines it (here we 
used that for the moment being we deal with smooth connections). Since the 
formula 
$\{A_a^i(y),V(x,\epsilon)\}\propto e_a^i(y)$ is always true provided that 
$\chi_\epsilon(x,y)=1$ we have that 
$\{h_e(t_k,t_{k+1}),V(e(t_k),\epsilon)\}\not=0$ whatever partition we 
choose. We may therefore choose for given $\gamma$ the partition, given
$\epsilon$, such that the following two conditions hold :\\
1) $\min_{e,k,a}(|e^a(t_k)-e^a(t_{k-1})|)>\epsilon$ and\\
2) $\min_{e\not=e',k+l>0,a}(|e^a(t_k)-e^{\prime a}(t_l)|)>\epsilon$.\\
Thus, the partition is as fine as we wish but only so fine that 
that $(t_k-t_{k+1})/\epsilon$ is at least of order $o(1)$.\\ 
We then can replace (\ref{5}) by 
\ba \label{6}
&&-\hat{H}^{E,\epsilon}_{YM}(N)f\nonumber\\
&=&\sum_{e,e'}\frac{\alpha_Q m_p}{2\ell_p^3}
\sum_{k,l=1}^n N(e(t_k))\chi_\epsilon(e(t_k),e'(t_l)) 
\times\nonumber\\&\times&  
\mbox{tr}(h_e(t_k,t_{k+1})[h_e(t_k,t_{k+1})^{-1},
\sqrt{\hat{V}(e(t_k),\epsilon)}]
\times\nonumber\\&\times&
h_{e'}(t_l,t_{l+1})[h_{e'}(t_l,t_{l+1})^{-1},
\sqrt{\hat{V}(e(t_l),\epsilon)}])
\times\nonumber\\
&\times&\{[
\mbox{tr}(\un{h}_e(0,t_k)\tau_I\un{h}_e(t_k,1)\partial/\partial\un{h}_e(0,1))
\mbox{tr}(\un{h}_{e'}(0,t_l)\tau_I\un{h}_{e'}(t_l,1)
\partial/\partial\un{h}_{e'}(0,1))]
\nonumber\\
&&+\delta_{e,e'}[\theta(t_l-t_k)
\mbox{tr}(\un{h}_e(0,t_k)\tau_I\un{h}_e(t_k,t_l)\tau_I\un{h}_e(t_l,1)
\partial/\partial\un{h}_e(0,1)) \nonumber\\
&&+\theta(t_k-t_l)
\mbox{tr}(\un{h}_e(0,t_l)\tau_I\un{h}_e(t_l,t_k)\tau_I\un{h}_e(t_k,1)
\partial/\partial\un{h}_e(0,1))]\}f.
\ea
Consider first the terms with $e\not=e'$. Then for sufficiently small 
$\epsilon$ we get 
$\chi_\epsilon(e(t_k),e'(t_l))=0$ unless $e,e'$ intersect each 
other. We have set up the problem in such a way that they then must
intersect in a vertex $e(0)=e'(0)=v$ of the graph $\gamma$. Now, by
condition 2) on our partition we also obtain that 
$\chi_\epsilon(e(t_k),e'(t_l))=0$ unless $t_k=t_l=0$.\\
If $e=e'$ then 
$\chi_\epsilon(e(t_k),e(t_l))=0$ unless $k=l$ by condition 1) on the 
partition. Now we make use of the 
fact that we can commute the gravitational operators with the Yang-Mills
operators. We then find out that 
\ba \label{7}
&&\mbox{tr}(
h_e(t_k,t_{k+1})[h_e(t_k,t_{k+1})^{-1},\sqrt{\hat{V}((e(t_k),\epsilon)}]
\times\nonumber\\&\times&  
h_{e}(t_k,t_{k+1})[h_{e}(t_k,t_{k+1})^{-1},
\sqrt{\hat{V}((e(t_k),\epsilon)}])f \nonumber\\
&=&
-\mbox{tr}(
[h_e(t_k,t_{k+1}),\sqrt{\hat{V}((e(t_k),\epsilon)}]
[h_{e}(t_k,t_{k+1})^{-1},
\sqrt{\hat{V}((e(t_k),\epsilon)}])f
\ea
vanishes unless $k=0$ because the volume operator $\hat{V}(x,\epsilon)$
annihilates a state unless there is a vertex in the region corresponding
to the $\epsilon-$box around $x$ and because, by definition, only the 
starting point of an edge is a vertex of the graph.
But by definition $\theta(0)=0$.
We thus conclude that for sufficiently small $\epsilon$ we obtain
\ba \label{8}
\hat{H}^\epsilon_{YM}(N)f
&=&-\frac{\alpha_Q m_p}{2\ell_p^3}\sum_{v\in V(\gamma)}\sum_{v\in e\cap e'}
\mbox{tr}(
h_e(0,\epsilon)[h_e(0,\epsilon)^{-1},\sqrt{\hat{V}(v,\epsilon)}]
\times\nonumber\\&\times&  
h_{e'}(0,\epsilon)[h_{e'}(0,\epsilon)^{-1},
\sqrt{\hat{V}(v,\epsilon)}])
\un{X}_e^I\un{X}_{e'}^I f
\ea
which does not depend on the details of the partition any longer because 
of which we could replace $t_1$ by $\epsilon$ ! Here we have defined the 
right-invariant vector fields 
$\un{X}^I(\un{g})=\mbox{tr}(\un{\tau}_I\un{g}\partial/\partial\un{g})$
and $\un{X}_e=\un{X}(\un{h}_e)$.
Notice that the final form of the operator
(\ref{8}) is manifestly gauge invariant. Also, we could actually replace
$\hat{V}(v,\epsilon)$ by $\hat{V}$ because the commutator $[h_e,\hat{V}]$
equals $[h_e,\hat{V}(e(0),R)]$ for any arbitrarily chosen neighbourhood
of $v=e(0)$, see \cite{5,6}. In particular, it equals 
$[h_e,\hat{V}_{e(0)}]$ where $\hat{V}_v$ denotes the volume operator at 
a point $v$. This operator is defined on any cylindrical function $f_\gamma$
as follows : consider an arbitrary finite contractable neighbourhood $R$ 
of $v$, denote by $R_t$ any homotopy with $R_1=R,\;R_0=\{v\}$  
and evaluate $\hat{V}(R_t)f_\gamma$. By the properties of the volume 
operator, the vector $\hat{V}(R_t)f_\gamma$ is constant for all $t<t_\gamma$ 
for some value $t_\gamma>0$ which depends on $\gamma$ whenever $R_t$ is 
so small that $v$ is possibly the only vertex of $\gamma$ contained in  
$R_t$ and the vector is 
moreover independent of $R$ and the homotopy. This vector is denoted by 
$\hat{V}_{v,\gamma} f_\gamma$. The family of operators 
$\{\hat{V}_{v,\gamma}\}_\gamma$ so defined is consistently defined because
$\hat{V}(R)$ is and therefore qualifies as the cylindrical projection 
of an operator $\hat{V}_v$ (see \cite{19a} for a different definition
in terms of germs of analytical edges). \\
This replacement of $\hat{V}(v,\epsilon)$ by $\hat{V}_v$ is possible only in 
the quantum version
where the square root of volume operator is defined via its spectral
resolution and automatically takes care of its local action while in the
classial computation we would have to keep the $\epsilon$. Thus the only
$\epsilon$ dependence of (\ref{8}) rests in the holonomies 
$h_e(0,\epsilon)$. But since the operator (\ref{8}) is gauge invariant,
by an argument given in \cite{16} even the remaining $\epsilon-$dependence
drops out as follows : we define for each edge $e$ of the graph incident 
at $v$ a segment $s(e)$
also starting at $v$ but not including the other endpoint of $e$. After
evaluating the operator on a state, the dependence on $s(e)$ automatically
drops out.\\
The final expression for the electrical part of the Yang-Mills Hamiltonian
is then given by 
\ba \label{9}
\hat{H}^E_{YM}(N)f
&=&-\frac{\alpha_Q m_p}{2\ell_p^3}\sum_{v\in V(\gamma)}
N_v \sum_{v\in e\cap e'}
\times\nonumber\\&\times&  
\mbox{tr}(h_{s(e)}[h_{s(e)}^{-1},\sqrt{\hat{V}_v}]
h_{s(e')}[h_{s(e')}^{-1},\sqrt{\hat{V}_v}])
\un{X}_e^I\un{X}_{e'}^I f
\ea
where $N_v=N(v)$.\\
Notice that we have exchanged the limits of taking $\epsilon\to 0$ and 
the limit of refining the partition ad infinitum. However, one could have 
arrived at (\ref{9}) also differently : let $\delta:=\inf_k (t_k-t_{k-1})$.
Make $\epsilon$ in (\ref{3}) $y$-dependent, that is,
$\epsilon(y)=\rho(y)\delta$ where $\rho(y)>0$ $d^3y$ almost everywhere
and such that the conditions 1),2) on the partition hold (with $\epsilon$
replaced by $\epsilon(y)$) at $y=e(t_k)$). Then instead 
of taking $\epsilon$ sufficiently small and the partion small but still 
finite we make $\epsilon$ dependent on $\delta$ in this sense and just 
take $\delta$ sufficiently small but still keep it finite. The result
(\ref{9}) is the same by construction, just that we did not need to take any 
limits and so
the questionable interchange of limiting procedures is unnecessary. Now,
since (\ref{9}) actually is independent of $\delta$, no limit needs to be 
taken. On the other hand, this latter regularization scheme is, in 
contrast to the former scheme, state-dependent although the final 
operator is state-independent as we will see in the next section.\\
\\
Let us now turn to the magnetic part. In this case we need to introduce a 
triangulation of $\Sigma$ just as in \cite{5,6} in order to define its
regularization. Taking over the notation from \cite{5,6} for the 
triangulation of $\Sigma$, for each vertex of 
$\gamma$ and each triple of edges $e,e',e^{\prime\prime}$ we introduce 
tetrahedra $\Delta$ with basepoint $v(\Delta)=v$
and incident segments $s_i(\Delta),\;i=1,2,3$ where there is a 
one to one map between the segments $s(e),s(e'),s(e^{\prime\prime})$ 
as defined above and the three $s_i(\Delta)$.
We have assumed that $\epsilon_{abc}\dot{s}_1^a
\dot{s}_2^b\dot{s}_3^c\ge 0$. We will denote the arcs of $\Delta$ that
connect the endpoints of $s_i(\Delta),s_j(\Delta)$ by
$a_{ij}(\Delta)$. Finally we have loops $\alpha_{ij}:=s_i\circ a_{ij}
\circ s_j^{-1}$.\\
Given first of all any triangulation of $\Sigma$
it is immediate to see that the magnetic part
of the Yang-Mills Hamiltonian can be written, by the same trick that we used 
for the electric part, as follows 
\ba \label{10}
H^B_{YM}(N)&=&\lim_{\epsilon\to 0}\frac{1}{2 Q^2}\sum_{\Delta,\Delta'}
\int_\Delta N(x) \{A^i(x),\sqrt{V(x,\epsilon)}\}\wedge\un{F}^I(x)
\times\nonumber\\&\times&  
\int_{\Delta'} \chi_\epsilon(x,y) 
\{A^i(y),\sqrt{V(y,\epsilon)}\}\wedge\un{F}^I(y).
\ea
Now notice that
\begin{eqnarray*}
&& f(v)\epsilon^{jkl}\mbox{tr}(\un{\tau}_I\un{h}_{\alpha_{jk}(\Delta)})
\times\nonumber\\&\times&  
\mbox{tr}(\tau_i h_{s_l(\Delta)}\{h_{s_l(\Delta)}^{-1},
\sqrt{V(v,\epsilon)}\})
\approx -2d6\int_\Delta f(x)
\{A^i(x),\sqrt{V(x,\epsilon)}\}\wedge\un{F}^I(x)
\end{eqnarray*}
where $f$ is any continuous function, $v=v(\Delta)$ and $\approx$ refers
to the expansion parameter $\delta$ of the left hand side in the 
parametrization of $s(e)=v+\delta \dot{s}(e)^a(0)+o(\delta^2)$, meaning that
the error is of order $\delta$. 
We now synchronize\footnote{This is justified because
the classical limit schemes $\lim_{\epsilon\to 0}\lim_{\delta\to 0}$ and 
$(\lim_{\delta\to 0})_{|\epsilon=\epsilon(\delta),\epsilon(0)=0}$ are 
identical, both give back the classical magnetic Hamiltonian constraint.
We will here neither give the details of the dependence $\epsilon(\delta)$ 
nor a proof but refer the reader to \cite{19a} for more details. We will 
take advantage of this fact without mentioning also in later subsections 
of this paper.} $\epsilon\approx\delta$  and it 
follows that we can write a regulated operator 
\ba \label{11}
\hat{H}^B_{YM,\epsilon}(N)&=&-\frac{m_p}{2\alpha_Q(12d)^2\ell_p^3}
\sum_{\Delta,\Delta'}N(v(\Delta))\epsilon^{jkl} 
\times\nonumber\\&\times&  
\mbox{tr}([\tau_i h_{s_l(\Delta)}[h_{s_l(\Delta)}^{-1},\sqrt{\hat{V}_v}])
\mbox{tr}(\un{\tau}_I\un{h}_{\alpha_{jk}(\Delta)})\times\nonumber\\
&\times& \chi_\epsilon(v(\Delta),v(\Delta')) \epsilon^{mnp}
\mbox{tr}([\tau_i h_{s_p(\Delta')}[h_{s_p(\Delta')}^{-1},\sqrt{\hat{V}_v}])
\times\nonumber\\&\times&  
\mbox{tr}(\un{\tau}_I\un{h}_{\alpha_{mn}(\Delta')})
\ea
where we could again drop the $\epsilon$ dependence in the argument of
the volume operator. The negative sign in (\ref{11}) stems from the 
$(-i)^2$ coming from replacing the Poisson brackets by commutators times
$1/(i\hbar)$.\\
So far everything was true for an arbitrary triangulation. We now apply the
operator (\ref{11}) to a function $f$ cylindrical with respect to a graph
$\gamma$ and adapt the triangulation to the graph in exactly the same way
as in \cite{5,6} and as indicated above. Let $E(v)=n(v)(n(v)-1)(n(v)-2)/6$
where $n(v)$ is the valence of the vertex $v$. As we evaluate the operator
we find out that only those terahedra $\Delta$ in (\ref{11}) contribute 
whose basepoint $v(\Delta)$ coincides with a vertex $v$ of the graph due to 
the presence of the volume operators in (\ref{11}). This mechanism 
is explained in more detail in \cite{5,6}. Moreover, as we take 
$\epsilon$ sufficiently small we see that only pairs of tetrahedra contribute 
which have the same 
basepoint $v(\Delta)=v(\Delta')$. Combining both observations, we find 
that we need to sum only over vertices of the graph and for each vertex $v$
over those tetrahedra $\Delta$ such that $v(\Delta)=v$. Then we find 
\ba \label{12}
&&\hat{H}^B_{YM,T}(N) f_\gamma=-\frac{m_p}{2\alpha_Q(12d)^2\ell_p^3}
\sum_{v\in V(\gamma)}N_v(\frac{8}{E(v)})^2\sum_{v(\Delta)=v(\Delta')=v}
\times\nonumber\\&\times&  
\epsilon^{jkl} 
\mbox{tr}([\tau_i h_{s_l(\Delta)}[h_{s_l(\Delta)}^{-1},\sqrt{\hat{V}_v}])
\mbox{tr}(\un{\tau}_I\un{h}_{\alpha_{jk}(\Delta)})\times\nonumber\\
&\times& \epsilon^{mnp}
\mbox{tr}([\tau_i h_{s_p(\Delta')}[h_{s_p(\Delta')}^{-1},\sqrt{\hat{V}_v}])
\mbox{tr}(\un{\tau}_I\un{h}_{\alpha_{mn}(\Delta')})f_\gamma.
\ea 
The label ``$T$" on the operator in the first line of (\ref{12}) is to 
indicate its dependence on the triangulation \cite{5,6} which 
expresses itself partly in the huge freedom of how to choose the loops 
$\alpha_{ij}$. This arbitrariness is somewhat reduced 
in the diffeomorphism invariant context that we are interested in because 
then it does not matter how ``large" the loops $\alpha_{ij}$ are as long
as the prescription how to attach them is diffeomorphism covariant. See
\cite{5,6} for further discussion of this point.\\
This furnishes the regularization of the Yang-Mills Hamiltonian constraint.

\subsection{Fermionic sector}

In this section we will only focus on the first term displayed in the 
expression for $H_{Dirac}$ in (\ref{1}). The other two terms can be quantized
similarily, for the quantization of $K_a^i$ we adopt a procedure identical
to the one used for the quantization of  
Einstein contribution to the Hamiltonian constraint in \cite{5,6}. This 
point is also outlined in appendix A.

We begin by rewriting the classical constraint using 
that $E^a_i=\frac{1}{2}\epsilon^{abc}\epsilon_{ijk}e_b^j e_c^k$. We find
by an already familiar procedure that
\ba \label{13}
&& H_{Dirac}(N)
\nonumber\\
&=&-\frac{i}{2\kappa^2}
\int d^3x N(x)\epsilon^{ijk}\epsilon^{abc}
\frac{4\{A_a^i(x),V(x,\delta)\}
\{A_b^j(x),V(x,\delta)\}}{\sqrt{\det(q)}(x)}
[(\tau_k{\cal D}_c\xi)_{A\mu}(x)\pi_{A\mu}(y)-c.c.]\nonumber\\
&&
\ea
where $\delta$ is an arbitrarily small bu finite parameter.
The minus sign comes from moving the classical momentum variable to the right
as compared to (\ref{1}).\\
The first task is to rewrite (\ref{13}) in terms of the quantities $\theta$.
To that end let $f^a_i$ be a real valued, $\mbox{ad}_{SU(2)}$ transforming
vector field and consider the discrete sum (we abbreviate $A\mu$ etc.
as $I$ etc.) \be \label{13a}
\sum_x f^a_i(x) (\tau_i{\cal D}_a\theta)_I(x)\bar{\theta}_I(x)\;.
\ee
Recall from \cite{TTKin} the definition $\theta_I(x):=\int d^3y 
\sqrt{\delta(x,y)}\xi_I(y):=\lim_{\epsilon\to 0} \theta^\epsilon_I(x)$
where $\theta^\epsilon_I(x)=\int d^3y 
\frac{\chi_\epsilon(x,y)}{\sqrt{\epsilon^3}}\xi_I(y)$ and 
$\chi_\epsilon(x,y)$
denotes the characteristic function of a box with Lebesgue measure 
$\epsilon^3$ and centre $x$. We define 
$(\partial_a\theta_I)(x):=\lim_{\epsilon\to 
0}\partial_{x^a}\theta^\epsilon(x)$ and find 
\ba
\partial_{x^a}\theta^\epsilon_I(x)&=&
\int d^3y 
\frac{\partial_{x^a}\chi_\epsilon(x,y)}{\sqrt{\epsilon^3}}\xi_I(y)
\nonumber\\
&=&-\int d^3y 
\frac{\partial_{y^a}\chi_\epsilon(x,y)}{\sqrt{\epsilon^3}}\xi_I(y)
=\int d^3y 
\frac{\chi_\epsilon(x,y)}{\sqrt{\epsilon^3}}\partial_{y^a}\xi_I(y)
\nonumber
\ea
since $\chi_\epsilon(x,y)=\chi_\epsilon(y,x)$ and there was no boundary 
term dropped in the integration by parts because $\chi_\epsilon$ is of 
compact support. Let us partition $\Sigma$
by a countable number of boxes $B_n$ of Lebesgue measure $\epsilon^3$ and 
center $x_n$ as in 
\cite{TTKin} and interprete (\ref{13a}) as the $\epsilon\to 0$ limit of 
\be \label{13b}
\sum_n f^a_i(x_n) (\tau_i{\cal D}_a\theta^\epsilon)_I(x_n)
\bar{\theta}^\epsilon_I(x_n)\;. 
\ee
Substituting for $\theta^\epsilon$ in terms of $\xi$, (\ref{13b}) becomes
\be \label{13c}
\int d^3x\int d^3y 
[\sum_n f^a_i(x_n)
\frac{\chi_\epsilon(x,x_n)\chi_\epsilon(y,x_n)}{\epsilon^3}]
[(\tau_i\partial_a\xi_I(x)+(\omega_a(x_n)\xi(x))_I]\bar{\xi}_I(y)\;.
\ee
We have not written the Christoffel connection in \ref{13c} which is needed
due to the density weight of $\xi$ because it drops out in the 
anti-symmetric sum $i[(.)-(.)^\star]=i[(.)-c.c]$ of (\ref{13}).
Now, as $\epsilon\to 0$ (the partition of $\Sigma$ becomes finer and 
finer) we can replace $\chi_\epsilon(x,x_n)$ by
$\delta(x,x_n)$ and $\chi_\epsilon(y,x_n)$ by $\delta_{x_n,y}$ and 
(\ref{13c}) becomes, upon performing the $x-$integral and the sum over $x_n$,
\be \label{13d}
\int d^3x f^a_i(x)
(\tau_i{\cal D}_a\xi_I)(x)\bar{\xi}_I(y)
\ee
which is precisely (\ref{13}) with the proper interpretation of $f^a_i$.
Expression (\ref{13d}) is written in a form that is well defined on the 
kinematical Hilbert space which consists of functions of $\theta$ rather 
than $\xi$.\\
Now, in quantizing expression (\ref{13a}) we keep the fermionic momenta 
to the right and replace $\bar{\theta}_{A\mu}(x)$ by 
$\hbar\partial/\partial\theta_{A\mu}$ which is the proper quantization 
rule for the $\theta$ variables \cite{TTKin}.
Also, we multiply nominator and dominator by $\delta^3$ and replace 
$\delta^3\sqrt{\det(q)}(x)$ by $V(x,\delta)$ in the denominator which by
the standard trick we can absorb into the Poisson bracket. Finally
we replace the Poisson bracket by a commutator times $1/(i\hbar)$.
Labelling the regulated operator with the parameter $\delta$, we find on a 
function $f_\gamma$ cylindrical with respect to a graph $\gamma$
with fermionic insertions $\theta_{A\mu}$ at the vertices $v\in V(\gamma)$
\ba \label{14}
&&\hat{H}^\delta_{Dirac}(N)f_\gamma=-\frac{\hbar}{2\ell_p^4}
\sum_{v\in V(\gamma)}\sum_x N(x)\epsilon^{ijk}
\epsilon^{abc}
\times\nonumber\\&\times&\delta^3  
[A_a^i(x),\sqrt{\hat{V}(x,\delta)}]
[A_b^j(x),\sqrt{\hat{V}(x,\delta)}]
[(\tau_k{\cal D}_c\theta)_{A\mu}(v)\frac{\partial}{\partial\theta_{A\mu}(v)}
\delta_{x,v}+h.c.]f_\gamma. 
\ea
Notice that the sum over all $x\in\Sigma$ already collapses to 
a sum over the vertices of $\gamma$.
Next we triangulate $\Sigma$ in adaption to $\gamma$.
We have the expansion
$H_s(0,\delta)\theta(s(\delta))-\theta(s(0))=\delta\dot{s}^a(0)({\cal
D}_a\theta)(s(0))$. Therefore we just introduce as in the sections before 
a holonomy at various places to absorb the factor of $\delta^3$ and 
replace $\hat{V}(v,\delta)$ by $\hat{V}_v$. Thus, 
\ba \label{15}
\hat{H}^\delta_{Dirac}(N)&=&-\frac{m_p}{2\ell_p^3}
\sum_{v\in V(\gamma)} N_v 
\times\nonumber\\&\times&  
\sum_{v(\Delta)=v} \epsilon^{ijk}\epsilon^{mnp}
\mbox{tr}(\tau_i h_{s_m(\Delta)}[h_{s_m(\Delta)}^{-1},\sqrt{\hat{V}_v}])
\times\nonumber\\ &\times& 
\mbox{tr}(\tau_j h_{s_n(\Delta)}[h_{s_n(\Delta)}^{-1},\sqrt{\hat{V}_v}])
\times\nonumber\\&\times&  
[(\tau_k[H_{s_p(\Delta)}\theta(s_p(\Delta)(\delta))-\theta(v)]_{A\mu}
\frac{\partial}{\partial\theta_{A\mu}(v)}+h.c.]
\nonumber\\
&=&-\frac{m_p}{2\ell_p^3}
\sum_{v\in V(\gamma)} N_v 
\sum_{v(\Delta)=v} \epsilon^{ijk}\epsilon^{mnp}\times\nonumber\\
&\times&
\mbox{tr}(\tau_i h_{s_m(\Delta)}[h_{s_m(\Delta)}^{-1},\sqrt{\hat{V}_v}])
\times\nonumber\\&\times&  
\mbox{tr}(\tau_j h_{s_n(\Delta)}[h_{s_n(\Delta)}^{-1},\sqrt{\hat{V}_v}])
[(Y_k(s_p(\Delta))-Y_k(v)+h.c.]
\nonumber\\
&=:& \hat{H}^T_{Dirac}
\ea
where the label $T$ reminds us of the triangulation dependence (we have 
naturally chosen the value of $\delta$ in such a way that a) $e(\delta)$
coincides with the endpoint of the segment of $e$ starting at $v=e(0)$
and b) is part of the definition of the triangulation adapted to $\gamma$).
We have defined
$$
Y_i(e):=\mbox{tr}(\tau_i H_e\xi(e(1))\frac{\partial}
{\partial\xi(e(0))})\mbox{ and }Y_i(v):=Y(e=v)
$$
and $e:[0,1]\to\Sigma$ is a suitable parametrization of the edge $e$.

The hermitian conjugation operation ``$h.c.$" involved in (\ref{15}) is 
meant with respect to the inner product on the Hilbert space and with 
respect to the operator of which the first term in (\ref{15}) is the 
projection on the cylindrical subspace labelled by th graph $\gamma$.
We will return to this issue in the next section.

Notice that the classical fermionic Hamiltonian constraint is a density
of weight one and that the operator defined by (\ref{15}) precisely respects 
this 
because the $\theta$ are scalar valued and not density-valued. If we were 
dealing with the $\xi$ instead of the $\theta$ we were running into conflict
with diffeomorphism covariance at this point.

\subsection{Higgs Sector}

We finally come to regularize the Higgs sector. Especially for this 
sector a general scheme will become evident of how to systematically
{\em take advantage} of the factor ordering ambiguity in order to arrive 
at a densely defined operator.

The term in (\ref{1}) proportional to $(p^I)^2$ looks hopelessly 
divergent : even if we would manage to replace the denominator by the 
volume operator we end up with a singular, not densely defined operator 
because the volume operator has a huge kernel. We need a new trick 
as follows :\\
We insert
the number $1=[\det(e_a^i)]^2/[\sqrt{\det(q)}]^2$ (one) into the kinetic 
term which apparantly makes the singularity even worse. However, 
consider the following regulated {\em four-fold} point-splitting of the 
kinematical term 
\ba \label{17}
&& H^\epsilon_{Higgs,kin}(N)\nonumber\\
&=&\frac{1}{2\kappa}
\int d^3x N(x)p^I(x)\int d^3y\; p^I(y)
\int d^3u (\frac{\det(e_a^i)}{[\sqrt{V(u,\epsilon)}]^3})(u)
\int d^3v (\frac{\det(e_a^i)}{[\sqrt{V(v,\epsilon)}]^3})(v)
\times\nonumber\\
&\times&\chi_\epsilon(x,y)\chi_\epsilon(u,x)\chi_\epsilon(v,y)\nonumber\\
&=&\frac{1}{2\kappa} \frac{(-2)^2}{(3!)^2\kappa^6}
\int d^3x N(x) p^I(x)\int d^3y p^I(y)\times\nonumber\\
&\times& \int
\mbox{tr}(\{A(u),\sqrt{V(u,\epsilon)}\}\wedge\{A(u),\sqrt{V(u,\epsilon)}\}
\wedge\{A(u),\sqrt{V(u,\epsilon)}\})\times\nonumber\\
&\times& \int
\mbox{tr}(\{A(v),\sqrt{V(v,\epsilon)}\}\wedge\{A(v),\sqrt{V(v,\epsilon)}\}
\wedge\{A(v),\sqrt{V(v,\epsilon)}\})\times\nonumber\\
&\times& \chi_\epsilon(x,y)\chi_\epsilon(u,x)\chi_\epsilon(v,y)\;.
\ea
Recall that $\int d^3x \det(e_a^i)=
\frac{1}{3!}\int \epsilon_{ijk}e^i\wedge e^j\wedge e^k
=-\frac{1}{3}\int \mbox{tr}(e\wedge e\wedge e)$ in order to see this. Notice 
that the sign factor in the identity (\ref{2}) has dropped out. We could 
also have used $1=\mbox{sgn}(\det(e_a^i))\det(e_ai)/\sqrt{\det(q)}$ but 
then the resulting expression would be less symmetric, it is a choice of 
factor ordering.\\ 
Now we replace 
$p^I$ by $-i\hbar(\kappa) \delta/\delta\phi^I$, replace the volume by
its operator version and Poisson brackets by commutators times 
$1/(i\hbar)$ and find, 
when applying the operator to a cylindrical function $f_\gamma$, that
\ba \label{18}
&& \hat{H}^\epsilon_{Higgs,kin}(N)f_\gamma\nonumber\\
&=&\frac{(-i)^2}{i^6}\frac{\hbar^2\kappa^2}{18\hbar^6\kappa^7}
\sum_{v,v'\in V(\gamma)} N(v) X^I(v)X^I(v')
\chi_\epsilon(v,v') \times\nonumber\\
&\times&  
\int
\mbox{tr}([A(x),\sqrt{\hat{V}(x,\epsilon)}]\wedge
[A(x),\sqrt{\hat{V}(x,\epsilon)}]\wedge[A(x),\sqrt{\hat{V}(x,\epsilon)}])
\times\nonumber\\
&\times& \int
\mbox{tr}([A(y),\sqrt{\hat{V}(y,\epsilon)}]\wedge
[A(y),\sqrt{\hat{V}(y,\epsilon)}]\wedge[A(y),\sqrt{\hat{V}(y,\epsilon)}])
f_\gamma
\chi_\epsilon(x,v)\chi_\epsilon(y,v')\;.
\nonumber\\
& &\ea 
where $X^I(v):=\frac{1}{2}[X^I_R(U(v))+X^I_L(U(v))]$ is the symmetric sum
of right and left invariant vector fields at $U(v)\in G$. 
The appearance of $X^I(v)$ relies on the following consideration, 
explained in more detail in \cite{TTKin} : Instead of $\hat{p}^I(x)$
we consider the integrated quantity $\hat{p}^I(B)$ where $B$ is a compact 
region in $\Sigma$. Now the functional derivative of $U(v)$ with 
respect to $\phi_I(x)$ turns out to be meaningless without a 
regularization of $U(v)$ as well. In \cite{TTKin} we use a regularization
which takes the interpretation of $U(v)$ as the nontrivial limit of a 
holonomy $\un{h}_e$ as $e$ shrinks to $e$ serious. Now the functional 
derivative can be unambiguously performed and since the functional derivative
of a holonomy of a connection along an edge gives rise to right or left 
invariant vector fields respectively at the beginning or end of the edge
respectively
it is not surprising that as we remove the regulator on $U(v)$ that we obtain
a symmetric sum of right and left invariant vector fields. The result is 
that then $\hat{p}^I(B)U(v)=-i\hbar\kappa \chi_B(v)X^i(v) U(v)$.

Certainly we are now going to triangulate $\Sigma$ in adaption to 
$\gamma$ in an already familiar fashion and write 
\ba \label{18a}
&& \int_\Delta
\mbox{tr}([A(x),\sqrt{\hat{V}(x,\epsilon)}]\wedge
[A(x),\sqrt{\hat{V}(x,\epsilon)}]\wedge[A(x),\sqrt{\hat{V}(x,\epsilon)}])
\nonumber\\
&\approx&\frac{1}{6}\epsilon^{ijk}
\mbox{tr}(h_{s_i(\Delta)}[h_{s_i(\Delta)}^{-1},
\sqrt{\hat{V}(v(\Delta),\epsilon)}]
\mbox{tr}(h_{s_j(\Delta)}[h_{s_j(\Delta)}^{-1},
\sqrt{\hat{V}(v(\Delta),\epsilon)}]\times\nonumber\\
&\times& 
\mbox{tr}(h_{s_k(\Delta)}[h_{s_k(\Delta)}^{-1},
\sqrt{\hat{V}(v(\Delta),\epsilon)}])
\ea
which results in 
\ba \label{18b}
&& \hat{H}^\epsilon_{Higgs,kin}(N)f_\gamma\nonumber\\
&=&\frac{m_p}{18\ell_p^9}\frac{1}{36}
\sum_{p,q,r,s\in V(\gamma)} N(p) 
X^I(p)X^I(q) \chi_\epsilon(p,q) \times\nonumber\\
&\times&  \frac{8}{ E(r)}\chi_\epsilon(r,p)\sum_{v(\Delta)=r}\epsilon^{ijk}
\frac{8}{ E(s)}\chi_\epsilon(s,q)\sum_{v(\Delta')=s}\epsilon^{lmn}
\times\nonumber\\
&\times&
\mbox{tr}(h_{s_i(\Delta)}[h_{s_i(\Delta)}^{-1},
\sqrt{\hat{V}(v(\Delta),\epsilon)}]
\mbox{tr}(h_{s_j(\Delta)}[h_{s_j(\Delta)}^{-1},
\sqrt{\hat{V}(v(\Delta),\epsilon)}]\times\nonumber\\
&\times& \mbox{tr}(h_{s_k(\Delta)}[h_{s_k(\Delta)}^{-1},
\sqrt{\hat{V}(v(\Delta),\epsilon)}])
\nonumber\\
&\times&  
\mbox{tr}(h_{s_l(\Delta')}[h_{s_l(\Delta')}^{-1},
\sqrt{\hat{V}(v(\Delta'),\epsilon)}]
\mbox{tr}(h_{s_m(\Delta')}[h_{s_m(\Delta')}^{-1},
\sqrt{\hat{V}(v(\Delta'),\epsilon)}]\times \nonumber\\
&\times&
\mbox{tr}(h_{s_n(\Delta')}[h_{s_n(\Delta')}^{-1},
\sqrt{\hat{V}(v(\Delta'),\epsilon)}]) f_\gamma
\ea 
since only tetrahedra based at vertices of $\gamma$ contribute in the 
sum $\int_\Sigma=\sum_\Delta \int_\Delta$.\\

Now we just take $\epsilon$ to zero, realize that only terms with
$v=p=q=r=s$ contribute and find that
\ba \label{19}
&&\hat{H}_{Higgs,kin}(N)f_\gamma
=\frac{8 m_p}{9^2\ell_p^9}\sum_{v\in V(\gamma)} N(v)
X^I(v) X^I(v)\frac{1}{E(v)^2}\sum_{v(\Delta)=v(\Delta')=v}\times
\nonumber\\
&\times& 
\epsilon^{ijk}
\mbox{tr}(h_{s_i(\Delta)}[h_{s_i(\Delta)}^{-1},
\sqrt{\hat{V}_v}]
\mbox{tr}(h_{s_j(\Delta)}[h_{s_j(\Delta)}^{-1},
\sqrt{\hat{V}_v}]
\mbox{tr}(h_{s_k(\Delta)}[h_{s_k(\Delta)}^{-1},
\sqrt{\hat{V}_v}])
\times\nonumber\\
&\times&
\epsilon^{lmn}
\mbox{tr}(h_{s_l(\Delta')}[h_{s_l(\Delta')}^{-1},
\sqrt{\hat{V}_v}]
\mbox{tr}(h_{s_m(\Delta')}[h_{s_m(\Delta')}^{-1},
\sqrt{\hat{V}_v}]
\mbox{tr}(h_{s_n(\Delta')}[h_{s_n(\Delta')}^{-1},
\sqrt{\hat{V}_v}]) f_\gamma\;.\nonumber\\
&&
\ea 
The operator (\ref{19}) is 
certainly quite complicated but it is densely defined !\\
Next we turn to the term containing the derivatives of the scalar field. 
We write 
$$q^{ab}\sqrt{\det(q)}=\frac{E^a_i E^b_i}{\sqrt{\det(q)}}\mbox{ and }
E^a_i= \epsilon^{acd}\epsilon_{ijk}\frac{e_c^j e_d^k}{2} 
$$ 
and regulate (again we could have chosen to replace only one of the 
$E^a_i$ by the term quadratic in $e_a^i$ and still would arrive at a
well-defined result at the price of losing symmetry of the expression)
\ba \label{20}
&&H^\epsilon_{Higgs,der}(N)\nonumber\\
&=&\frac{1}{2\kappa}
\int d^3x\int d^3y N(x)\chi_\epsilon(x,y) \epsilon^{ijk}\epsilon^{imn}
\epsilon^{abc}
\frac{({\cal D}_a\phi_I e_b^j e_c^k)(x)}{\sqrt{V(x,\epsilon)}}
\epsilon^{bef}
\frac{({\cal D}_b\phi_I e_e^m e_f^n)(y)}{\sqrt{V(y,\epsilon)}}
\nonumber\\
&=&\frac{1}{2\kappa^5} (\frac{2}{3})^4
\int N(x) \epsilon^{ijk}
{\cal D}\phi_I(x)\wedge \{A^j(x),V(x,\epsilon)^{3/4}\} \wedge 
\{A^k(x),V(x,\epsilon)^{3/4}\}\times\nonumber\\
&\times& 
\int \chi_\epsilon(x,y) \epsilon^{imn}
{\cal D}\phi_I(y)\wedge \{A^m(x),V(y,\epsilon)^{3/4}\} \wedge 
\{A^n(y),V(y,\epsilon)^{3/4}\}\;.
\ea
It is clear where we are driving at. We replace Poisson brackets by
commutators times $1/i\hbar$ and $V$ by its operator version. Furthermore
we introduce the already familiar triangulation of $\Sigma$ and have, 
using that with $v=s(0)$ for some path $s$
\ba \label{20a}
&&\mbox{Ad}(\un{h}_s(0,\delta t))[U(s(\delta t))]-U(v)=
\un{h}_s(0,\delta t)U(s(\delta t))\un{h}_s(0,\delta t)^{-1}-U(v) 
\nonumber\\
&=&\exp(\un{h}_s(0,\delta t)\phi(s(\delta t))\un{h}_s(0,\delta t)^{-1})-U(v)
\nonumber\\
&=&\exp([1+\delta t\dot{s}^a(0)\un{A}_a][\phi(v)+\delta t\dot{s}^a(0)
\partial_a\phi(v)][1-\delta t\dot{s}^a(0)\un{A}_a]+o((\delta t)^2))-U(v)
\nonumber\\
&=&\exp(\delta t\dot{s}^a(0)(\partial_a\phi(v)+
[\un{A}_a,\phi(v)])+o((\delta t)^2))-U(v)=
\delta t\dot{s}^a(0){\cal D}_a\phi(v)+o((\delta t)^2),\nonumber\\
&&
\ea
and with 
$\mbox{tr}(\tau_i\tau_j)=-\delta_{ij}/2,\mbox{tr}(\un{\tau}_I\un{\tau}_J)
=-d\delta_{IJ}$, $d$ the dimension of the fundamental representation of $G$ 
that 
\ba \label{20b}
& &6\int_\Delta 
{\cal D}\phi_I(x)\wedge \{A^j(x),V(x,\epsilon)^{3/4}\} \wedge 
\{A^k(x),V(x,\epsilon)^{3/4}\}\nonumber\\
&\approx& -\frac{4}{d}\epsilon^{mnp}
\mbox{tr}(\un{\tau}_I[\mbox{Ad}(\un{h}_{s_m(\Delta)})[U(s_m(\Delta))]
-U(v(\Delta))])\times\nonumber\\
&\times& \mbox{tr}(\tau_j h_{s_n(\Delta)}\{h_{s_n(\Delta)}^{-1},
V(v(\Delta),\epsilon)^{3/4}\})
\mbox{tr}(\tau_k h_{s_p(\Delta)}\{h_{s_p(\Delta)}^{-1},
V(v(\Delta),\epsilon)^{3/4}\})\;.
\ea
Then we find on a cylindrical function 
\ba \label{21}
&&\hat{H}^\epsilon_{Higgs,der}(N)f_\gamma=
\frac{1}{2\kappa^5\hbar^4} (\frac{2}{3})^4 (\frac{2}{3d})^2
\sum_{v,v'\in V(\gamma)} N(v) \chi_\epsilon(v,v')
\epsilon^{ijk}\epsilon^{ilm}\times\nonumber\\
&\times& \sum_{v(\Delta)=v} \frac{8}{E(v)}\epsilon^{npq}
\mbox{tr}(\un{\tau}_I[\mbox{Ad}(\un{h}_{s_n(\Delta)})[U(s_n(\Delta))]
-U(v(\Delta))])\times\nonumber\\
&\times& \mbox{tr}(\tau_j h_{s_p(\Delta)}[h_{s_p(\Delta)}^{-1},
\hat{V}_v^{3/4}])
\mbox{tr}(\tau_k h_{s_q(\Delta)}[h_{s_q(\Delta)}^{-1},
\hat{V}_v^{3/4}])\times\nonumber\\
&\times& \sum_{v(\Delta')=v'} \frac{8}{E(v')}\epsilon^{rst}
\mbox{tr}(\un{\tau}_I[\mbox{Ad}(\un{h}_{s_r(\Delta')})[U(s_r(\Delta'))]
-U(v(\Delta'))])\times\nonumber\\
&\times& \mbox{tr}(\tau_l h_{s_s(\Delta')}[h_{s_s(\Delta')}^{-1},
\hat{V}_{v'}^{3/4}])
\mbox{tr}(\tau_m h_{s_t(\Delta')}[h_{s_t(\Delta')}^{-1},
\hat{V}_{v'}^{3/4}])f_\gamma
\ea
since only tetrahedra with vertices as basepoints contribute. Thus we find
in the limit $\epsilon\to 0$
\ba \label{22}
&&\hat{H}_{Higgs,der}(N)f_\gamma=
\frac{4^6 m_p}{2\ell_p^9 d^2 3^6} 
\sum_{v\in V(\gamma)} N(v) 
\epsilon^{ijk}\epsilon^{ilm}\times\nonumber\\
&\times& \sum_{v(\Delta)=v(\Delta')=v} \frac{1}{E(v)^2}
\epsilon^{npq}\epsilon^{rst}
\mbox{tr}(\un{\tau}_I[\mbox{Ad}(\un{h}_{s_n(\Delta)})[U(s_n(\Delta))]
-U(v)])\times\nonumber\\
&\times& \mbox{tr}(\tau_j h_{s_p(\Delta)}[h_{s_p(\Delta)}^{-1},
\hat{V}_v^{3/4}])
\mbox{tr}(\tau_k h_{s_q(\Delta)}[h_{s_q(\Delta)}^{-1},
\hat{V}_v^{3/4}])\times\nonumber\\
&\times& 
\mbox{tr}(\un{\tau}_I[\mbox{Ad}(\un{h}_{s_r(\Delta')})[U(s_r(\Delta'))]
-U(v)])\times\nonumber\\
&\times& \mbox{tr}(\tau_l h_{s_s(\Delta')}[h_{s_s(\Delta')}^{-1},
\hat{V}_v^{3/4}])
\mbox{tr}(\tau_m h_{s_t(\Delta')}[h_{s_t(\Delta')}^{-1},
\hat{V}_v^{3/4}])f_\gamma\;.
\ea
Again, despite its complicated appearence, (\ref{22}) defines a densely 
defined operator.
Finally the potential term, like the cosmological constant term are trivial
to quantize. Notice that certain functions of $\phi_I(v)\phi_I(v)$ can be 
recovered 
from polynomials of the functions $[\mbox{tr}(U(v)^n)]^m$ where $m,n$ are
non-negative integers. For instance for $SU(2)$ we have 
$2\cos(\sqrt{\phi_i(v)^2 })=\mbox{tr}(U(v))$. Thus we may {\em define}
for instance a mass term through 
$$
\phi_i(v)^2:=[\mbox{arcos}(\frac{\mbox{tr}(U(v))}{2})]^2
$$
where the arcos-function is for the principal branch and is well-defined 
because the argument takes values in $[-1,1]$ only. Thus, by this rule
all polynomials in $\phi_I(v)^2$ become actually bounded functions of 
$U(v)$. This is not an unknown phenomenon, the same happens when one replaces
the Yang-Mills action by its regularized Wilson action on a fixed lattice
(our lattice, the triangulation, is not fixed, it ``floats" with the state).
Therefore, taking $\phi_I(v)^2$ as expressed through those products of 
traces we find
\ba \label{22a}
\hat{H}_{Higgs,pot}(N)f_\gamma&=&\frac{m_p}{\ell_p^3}
\sum_{v\in V(\gamma)}N_v
P(\phi_I\phi_I)(v)\hat{V}_v f_\gamma
\nonumber\\
\hat{H}_{cosmo}(N)f_\gamma&=&\frac{m_p\lambda}{\ell_p^3}
\sum_{v\in V(\gamma)}N_v \hat{V}_v
f_\gamma\;.
\ea
This furnishes the quantization of the matter sector. Notice that all
Hamiltonians have the same structure, namely an operator which carries
out a discrete operation on a cylindrical function, like adding or
subtracting lines, fermions or Higgs fields, multiplied by the Planck
mass and devided by an appropriate power of the Planck length which 
compensates the power of the Planck length coming from the action of the 
volume operator. 
It follows that in this sense the matter Hamiltonians are quantized in 
multipla of the Planck mass when we go to the diffeomorphism invariant
sector.

\subsection{A general regularization scheme}

In this subsection we describe a recipe by means of which a fairly
large calss of Hamiltonian densities of weight one which are diffeomorphism 
covariant and coupled to gravity can be turned into densely defined and, 
as we will see later, anomaly-free operators on the Hilbert space that we 
have defined. The 
resulting expression {\em does} suffer from a factor ordering {\em 
ambiguity} but {\em not} from a factor ordering {\em singularity}. 

The restrictions on the Hamiltonian density are as follows :\\
a) The matter canonical 
momenta $P$ of the theory are scalar densities of weight one and 
matter configuration variables are scalars (they may transform
non-trivially under $SU(2)\times G$). In case
that matter is a priori described by tensors, turn them into internal 
$SU(2)$ tensors by means of the triad and co-triad, the corresponding 
canonical transformation will add to the gravitational connection 
a piece $\kappa_a^i$ which is a real valued one-form and transforms 
homogenuously under $SU(2)$; thus the reality of $A_a^i$ is preserved 
under this canonical transformation. Other cases require a 
special treatment (for instance the case of the fermion fields). \\
b) Furthermore, it is assumed that all covariant derivatives 
are with respect to $A_a^i,\un{A}_a^I$, act only on configuration scalars 
$Q$ and are of first order only so 
that no Christoffel connection is needed. In case that the covariant
derivative is a priori given in terms of the spin connection and/or
acts on a tensor, write the tensor as before in terms of (co)triads and
the canonical configuration scalars. If the covariant derivative is
${\cal D}_a$ with respect to $A_a$ we just use that ${\cal D}_a e_b^i
=\epsilon^{ijk} K_b^j e_c^k$. If it is $D_a$ with respect to
the spin connection, use that $e_a^i$ is annihilated by $D_a$ and 
that $({\cal D}_a-D_a)v_{i..j}=K_a^k[\epsilon_{ikl}v_{l..j}+..+\epsilon_{jkl}
v_{i..l}]$. We see that all covariant derivatives can be cast into the 
desired form up to underived factors of $K_a^i$ which we write
as a Poisson bracket $\{A_a^i,\{H^E(1),V\}\}$ where $H^E$ is the Euclidean
Hamiltonian constraint \cite{5,6}. \\
Restrictions a),b) are just in order to state the theorem below 
in a compact form. The case of higher derivatives (as they actually occur 
in the Riemann curvature) just require a case by case analysis.

Consider then a general Hamiltonian density which is local and consists of
monomials of the form (we suppress all indices and contractions other than
tensor contractions) 
\be \label{S1} H_{m,n}(x)=[P(x)]^n
E^{a_1}(x)..E^{a_m}(x)
f_{m,n}[Q]_{a_1..a_m}(x)\frac{1}{[\sqrt{\det(q)}(x)]^{n+m-1}}.  
\ee 
Here $f_{m,n}[Q]$ is a  tensor of
density weight zero which is independent of $e_a^i,P$ and is a polynomial
consisting of sums of terms involving covariant derivatives of $Q$ of
first order and underived factors of $K_a^i$ or $F_{ab}^i$ such that the 
total number of their covariant indices is $m$. The
denominator accounts for the fact that the Hamiltonian is a density of
weight one.\\ 
Expression (\ref{S1}) defines the most general basic building block
of the Hamiltonians under consideration, that is, every Hamiltonian in the
class that we have defined is a linear combination of these.
\begin{Theorem}[Structure Theorem] Any Hamiltonian constraint of the form
(\ref{S1}) can be turned into a densely defined operator on $\cal H$ which
is diffeomorphism covariantly defined and anomaly free. 
\end{Theorem}
Proof :\\ 
In case $m+n=0$ we are done because upon quantization
we just need to triangulate $\Sigma$ and replace $\sqrt{\det(q)(x)}$ by
$\hat{V}_x$.\\ 
Consider then the case $m+n>0$. In order to regulate
(\ref{S1}) we will need $m+n-1$ point splittings for the $n$ momenta $P,E$.
This will require $m+n-1$ regulated $\delta$-distributions
$\chi_\epsilon/\epsilon^3$.  Each factor factor $1/\epsilon^3$ can be
absorbed by replacing $1/\sqrt{\det(q)(x)}$ by $1/V(x,\epsilon)$ but then
we cannot simply replace this by its operator version because
$V(x,\epsilon)$ is in the denominator. \\ 
Now multiply (\ref{S1}) by the
number $1=[|\det(e_a^i)|/\sqrt{\det(q)}]^k$ and introduce $k>0$ more point
splittings. We have a power of $3k$ co-triads $e_a^i$ in the nominator
and a power of $n+m+k-1$ factors of $\epsilon^3\sqrt{\det(q)}$ in the 
denominator.
We replace each $\epsilon^3\sqrt{\det(q)(x)}$ by $V(x,\epsilon)$ following
the standard trick.\\ 
Now let $e:=\mbox{sgn}(\det(e_a^i))=(e)^3$. We have the
classical identity, using (\ref{2}) 
\ba \label{S2} 
&&|\det(e_a^i)(x)|=
\frac{1}{3!}e(x)\epsilon^{abc}\epsilon_{ijk}(e_a^i e_b^j e_c^k)(x)
=\frac{1}{3!}\epsilon^{abc}\epsilon_{ijk}[(e e_a^i)(e
e_b^j)(e e_c^k)](x) 
\nonumber\\ 
&=&
\frac{1}{3!(2\kappa)^3}\epsilon^{abc}\epsilon_{ijk}
\{A_a^i(x),V(x,\epsilon)\}\{A_b^j(x),V(x,\epsilon)\}
\{A_c^k(x),V(x,\epsilon)\} 
\ea 
and therefore each $e_a^i$ is worth a
factor of $V(x,\epsilon)$ in the nominator within a Poisson bracket. Now
choose $k$ large enough until $3k>n+m+k-1$, i.e. $2k>n+m-1$. By suitably
point splitting the various factors we get $3k$ factors of the form 
$$
\frac{\{A_a^i(x),V(x,\epsilon)\}}{V(x,\epsilon)^{\frac{n+m+k-1}{3k}}}
=\frac{\{A_a^i(x),V(x,\epsilon)^{1-\frac{n+m+k-1}{3k}}\}}{1-\frac{n+m+k-1}{3k}}
$$ 
and thus have managed to produce a net positive power of volume
functionals for the point-split classical Hamiltonian density in each of 
the $3k$ Poisson brackets. We now
choose the arguments $x,y$ of the various $\chi_\epsilon(x,y)$ so that the
limit $\epsilon\to 0$ gives a non-vanishing result only if all arguments
coincide. We triangulate $\Sigma$ in the fashion outlined in the previous
subsections and replace $A$ in
$\int_\Delta\mbox{tr}(\{A,V\}\wedge\{A,V\}\wedge\{A,V\})$ by the
holonomies along the edges of the triangulation. Finally we replace $P,E$,
ordered to the right, by the corresponding functional derivatives, Poisson
brackets by commutators and the volume functional by its operator version.
The result when applied at finite $\epsilon$ to a function cylindrical
with respect to a graph only gives contributions at an $m+n+3k$ tupel of
vertices (or edges) of the graph and when sending $\epsilon$ to zero all 
vertices of
the tupel have to coincide in order to give a non-vanishing result. This
shows that we find a densely defined operator. 

To see that it is anomaly-free we just note that the resulting operator is 
of the type to which the theorem of the section on anomaly-freeness applies.
\\
$\Box$\\
\\
Remarks :

1) We note that the density weight of one was crucial (besides the
fact that the integrated operator is only diffeomorphism invariant if the 
density weight is one) : \\
If it would have been higher than one then we needed $n+m+k-1$ point 
splittings but
in the denominator we have a power of $\sqrt{\det(q)}$ which is smaller than
$n+m+k-1$ and therefore even the regulated operator blows up at least
as $1/\epsilon^3$. If it was less than one then by a similar argument the 
regulated operator vanishes at least as $\epsilon^3$ which is trivially 
always zero.

2) The proof shows precisely the sources of the factor ordering ambiguity :\\
a) That we chose the momenta to the right was essentially forced on us 
because we want to obtain a densely defined operator : if the functional 
derivatives act on $f_{m,n}(Q)$ then in general it will not be true any 
longer that the operator only acts at the vertices of the state but at all
vertices of the triangulation which are infinite in number and so the 
resulting state would not be normalizable.\\
b) As long as $2k>n+m-1$ we can have arbitrarily large $k$ and still get 
a well-defined result. Surely, minimal $k$ is desirable to obtain a 
simple result.\\
c) We could have absorbed different powers of $V(x,\epsilon)$ into the 
various Poisson brackets, however, all powers must add up to $n+m+k-1$.\\
d) We are 
free to take advantage of the classical identity \be \label{S3}
E^a_i(x)=\frac{1}{2}\epsilon^{abc}\epsilon_{ijk}e_b^j(x) e_c^k(x)
=\frac{1}{8\kappa^2}\epsilon^{abc}\epsilon_{ijk}
\{A_b^j(x),V(x,\epsilon)\} \{A_c(x)^k,V(x,\epsilon)\} 
\ee 
to lower the necessary value of $k$ if desirable since each of the 
$m$ factors of $E$ is worth a power of two of $V(x,\epsilon)$. Of course,
it may be true that the value of $k$ must be at least one in order to
get a densely defined operator (if $k>0$ then for sure the resulting 
operator will act only at vertices of the graph as we proved).

3) The theorem works the same way in any dimension $d\ge 2$ because the 
critical
condition $dk>n+m+k-1$ can be satisfied by some $k$ for any value of $n,m$ 
only for $d>1$. However, in one spatial dimension all tensors are 
densities and in zero spatial dimension we do not have a field 
theory so that the theorem does not apply in these cases anyway.\\

\section{Consistency}

In this section we will perform the required consistency checks necessary to
show that we really constructed covariantly defined, anomaly-free, linear
operators through their action on cylindrical functions which is 
non-trivial in the sense that it has a non-vanishing kernel.

\subsection{Cylindrical Consistency}

Notice that, just like the gravitational Hamiltonian constraint, the 
matter Hamiltonian constraints are actually defined as a family of 
operators $\{\hat{H}_I\}$ where $I$ is a compound label consisting of the 
graph, colours and spins of its edges and fermionic and 
Higgs representations of its vertices, that is, 
$I=(\gamma,[\vec{j},\vec{n},\vec{p}],[\vec{c},\vec{C},\vec{q}])
=:(\gamma,\vec{\lambda})$. 
The set of 
labels $I$ is an uncountably infinite one because the set of piecewise 
analytical graphs of $\Sigma$  has this cardinality. Still this set 
allows for a nice and controllable orthogonal decomposition of the 
Hilbert space. In analogy with the source-free case, as the reader 
can easily prove himself given the measures defined in section 2, we
have : 
\be \label{23} 
{\cal H}=\oplus_\gamma {\cal H}_\gamma,\;
{\cal H}_\gamma=\oplus_{\vec{\lambda}} {\cal H}_{\gamma,\vec{\lambda}}
\ee
where the first direct sum is an uncountable one while the second 
is a countable one, running over the possible colourings of the graph 
with the various compatible irreducible representations, compatible in 
the sense that there exist projectors which render the associated 
cylindrical functions into gauge-invariant ones. The Hilbert space
${\cal H}_\gamma$ is infinite dimensional and is the completion of the 
space of functions built from spin-colour-network states on $\gamma$
while ${\cal H}_{\gamma,\vec{\lambda}}$ is a finite dimensional vector 
space with dimension equal to the number of linearly independent 
projectors on gauge invariant functions compatible with the colouring
$\vec{\lambda}$ of $\gamma$.\\ 
As the decomposition in (\ref{23}) is direct there exist orthogonal
projections $\hat{p}_{\gamma,\vec{\lambda}}\:\; {\cal H}\mapsto 
{\cal H}_{\gamma,\vec{\lambda}}$.
Cylindrical consistency now means that the family of operators 
$\{\hat{H}_{\gamma,\vec{\lambda}}\}$ is a family of projections of a 
single operator $\hat{H}$ defined on ${\cal H}$ such that 
$\hat{H}\hat{p}_{\gamma,\vec{\lambda}}=\hat{H}_{\gamma,\vec{\lambda}}$.
The necessary and sufficient condition for this to be the case is that
$\hat{H}_{\gamma,\vec{\lambda}}\hat{p}_{\gamma',\vec{\lambda}'}=0$ whenever
$\gamma\not=\gamma'$ or $\vec{\lambda}\not=\vec{\lambda}'$. But this is 
the case by construction if we simply define 
$\hat{H}:=\sum_{\gamma,\vec{\lambda}}\hat{H}_{\gamma,\vec{\lambda}} 
\hat{p}_{\gamma,\vec{\lambda}}$. This suffices to prove consistency.

\subsection{Diffeomorphism-Covariance, Continuum Limit\\
Self-Adjointness and Positive Semi-Definiteness}

We begin with diffeomorphism covariance of the family of operators 
obtained and the final 
continuum limit given by the infinite refinement of the triangulation.\\
Let $\Phi$ be the topological vector space constructed in 
\cite{4,TTKin} of finite linear combinations of 
spin-colour-network functions. By $\Phi'$ we mean the continuous linear
functionals on $\Phi$ and denote by $\hat{H}_\delta(N)$ the 
regulated Hamiltonian constraint 
where the parameter $\delta>0$ is a regularization parameter expressing 
the fact that the loops attached are finite in size, that is, we did not 
take the continuum limit yet. Let $f=\sum_I c_I T_I\in \Phi$ be a 
cylindrical function where $T_I$ are the spin-colour-network states.
As shown in \cite{4,TTKin}, the following 
object makes sense as a distribution on $\Phi$ : $[T_I]:=\sum_{T\in\{T_I\}} 
T$, where 
$\{T_I\}:=\{\hat{U}(\varphi)T_I\;:\; \varphi\in\mbox{Diff}(\Sigma)\}$ is 
the orbit of the vector $T_I$ under the diffeomorphism group. Moreover, 
as one can show, the $[T_I]$ provide an 
orthonormal basis on the diffeomorphism invariant Hilbert space, that is,
any $\Psi$ as above is a linear combination of those. Furthermore, we have 
$[f]:=\sum_I c_I [T_I]$, that is, every $f\in\Phi$ gets averaged term-wise
after decomposing it into spin-colour-network states (the reason for this
term-wise averaging is explained in \cite{4,TTKin}, we also neglected here
some technical details which one can also find in those papers). 
The definition of the inner product on the space of diffeomorphism invariant
distributions is given by  $<f,g>_{Diff}:=[f](g)$ where the latter 
expression means the evaluation of the distribution $[f]$ on the 
test function $g$ \cite{4,TTKin}. \\
A diffeomorphism invariant distribution $\Psi\in\Phi'$ is a solution of the 
Hamiltonian constraint provided that 
\be \label{23a}
\Psi(\hat{H}_\delta(N)f)=0
\ee
for any $f\in\Phi$ and $N\in{\cal S}$ (the space of test functions of rapid 
decrease). The striking feature of (\ref{23a}) is that it is {\it independent
of the value of $\delta$} ! The underlying reason for this
is the diffeomorphism covariance of the graph-dependent triangulation 
prescription. This is proved in \cite{6,18} for the 
gravitational part of the Hamiltonian constraint and the same reasoning 
applies to the matter coupled case as well and will not be repeated here.\\
The limit $\delta\to 0$ is therefore already performed in (\ref{23a})
(see also \cite{6}). Notice that, just as in the continuum limit of a 
regulated Euclidean path integral of constructive quantum field theory, 
we take the 
limit {\em after} integrating. Indeed, the limit before integrating does not
exist in the $L_2$ sense (for the same reason that the generator of the 
diffeomorphism constraint does not exist \cite{4}). \\
\\
Next we turn to the following issue :\\
If we set $N=1$ and the spatial metric has signature $(+,+,+)$ then 
the matter contribution to the integrated Hamiltonian {\em constraint}, 
at least for the Yang-Mills and the Higgs part, is classically a 
manifestly non-negative and diffeomorphism invariant functional.
Since upon replacing the dynamical gravitational field by some classical
background field this functional plays the role of the matter {\em 
Hamiltonian} in that background field, a natural question to ask 
is whether it should not be promoted to a positive-semi-definite, 
diffeomorphism invariant operator in the quantum theory (with dynamical
quantum gravitational field). In fact, this question is even more natural 
to ask in view of some kind of ``quantum dominant energy condition" on 
the Hamiltonian matter density (which equals the Hamiltonian constraint)
because it is the energy density component of the energy momentum tensor
(see \cite{19} for a first attempt towards a quantum formulation of a 
dominant energy condition in the quantum theory). \\
While imposing positivity is then very natural from this point of view it 
is, in fact, unnatural from another point of view : namely, if the 
matter Hamiltonian operator is positive semi-definite, then it is hard to
imagine how that can be true if not the Hamiltonian density, when 
integrated over any compact region of $\Sigma$, also becomes a positive 
semi-definite operator. In particular, the matter Hamiltonian constraint
when integrated against a non-negative lapse function should also be a 
postive 
semi-definite operator. However, then this latter operator will be 
automatically symmetric (it even would have self-adjoint extensions, at 
least its Friedichs extension). Now, by arguments explained in \cite{17}, 
a symmetric Hamiltonian constraint operator is in danger of being in 
conflict with the task of faithfully implementing the Dirac constraint 
algebra. Most certainly,
one expects a quantum anomaly in this case. Therefore, if one wants 
an anomaly-free quantum constraint algebra then it seems that one should not
insist on a positive semi-definite Hamiltonian constraint operator. One might
think that this can be accomplished, while keeping the matter Hamiltonian 
constraint positive, by having a non-symmetric gravitational Hamiltonian 
constraint but, at least on solutions, the gravitational and mater 
contribution just equal each other up to a sign and therefore necessarily
the gravitational piece is also symmetric if the matter piece is.

We will leave the resolution of this puzzle for future investigations and 
just mention that :\\
1) As we will see in the next section, the Dirac algebra is indeed not 
faithfully represented (one could, however, use the arguments of  
\cite{19a} to improve this), but still the algebra is 
non-anomalous in the sense that we obtain a consistent quantum theory,\\
2) The full matter Hamiltonian operator, as it stands in this paper, is 
not symmetric but at least it is diffeomorphism invariant,\\
3) The electric piece of the Yang-Mills Hamiltonian and the term bi-linear
in the Higgs momenta are indeed essentially self-adjoint and positive 
semi-definite operators on the diffeomorphism-invariant Hilbert space.\\
In the sequel demonstrate 2) and 3).\\
\\
Denote by $\hat{H}^m_\delta:=\hat{H}^m_\delta(1)$ the matter Hamiltonian 
constraint evaluated at unit lapse. 
We define the 
diffeomorphism invariant analogue of $\hat{H}^m_\delta$ in the continuum by
\be \label{23b}
<[T_I],\hat{H}^m[T_J]>_{Diff}:=\lim_{\delta\to 0+}[T_I](\hat{H}^m_\delta T_J)
\ee
where on the right hand side we have the evaluation of a distribution on 
a test function. 
Now, since 
$\hat{H}^m_\delta(N) T_J\in\Phi$ we have by definition
$[T_I](\hat{H}^m_\delta T_J)=<[T_I],[\hat{H}^m_\delta T_J]>_{Diff}$ but since
$[T_I]$ is diffeomorphism invariant, this number, as explained in 
(\ref{23a}), does not depend on
$\delta>0$ (the size or shape of the loop attached) and so is a constant.
The limit is therefore trivial and so
\be \label{23c}
<[T_I],\hat{H}^m[T_J]>_{Diff}=[T_I](\hat{H}^m_\delta T_J)
\ee
for any $\delta>0$. This displays $\hat{H}^m$ as a diffeomorphism invariant
operator.\\
\\
We will now demonstrate that (\ref{23c}) is not even symmetric :\\
We restrict ourselves to the case that we couple only a pure Yang-Mills
field. We just need 
one counter-example : Let $T_J$ be a spin-network state based on a graph 
$\gamma$ with three edges and two tri-valent vertices only one of which, $v$,
is such that the tangents of the edges are linearly independent at it. Colour 
the edges with suitably high irreducible representations of $SU(2)\times G$
such that $\hat{H}_{YM,\delta}^B T_J$ is a sum of spin-colour-network 
states $T_{J'}$ each of which depends on an extended graph $\gamma'$ of which
$\gamma$ is a proper subset, that is, $\gamma'=\gamma\cup\Delta$ where
$\Delta$ is a tetrahedron at $v$ defined by the triangulation. 
Now choose $T_I:=T_{J'}$ for one of the $J'$ so that 
$[T_{J'}](\hat{H}_{YM,\delta}^B T_J)\not=0$. On the other hand, if we apply
$\hat{H}_{YM,\delta}^B$ to $T_{J'}$ then we get a linear combination of
spin-network states which depend on graphs each of which 
includes $\gamma$ but is even larger than the graph $\gamma'$ 
which underlies
$J'$. It follows that $[T_J](\hat{H}_{YM,\delta}^B T_{J'})=0$ thus 
contradicting symmetry. A similar argument reveals that the Dirac 
and Higgs Hamiltonians as defined by (\ref{23b}) cannot be symmetric.\\
\\
Let us then conclude this subsection with showing that the 
electric piece of the Yang-Mills 
Hamiltonian is a densely defined, positive definite essentially self-adjoint 
operator even on the auxiliary Hilbert space $\cal H$. A similar argument
applies to the $p^I p^I$ part of the Higgs Hamiltonian.\\
Let $\hat{O}_s^{AB}:=(h_s[h_s^{-1},\hat{V}])_{AB}$ for any segment $s$
where $A,B$ denote $SU(2)$
indices. Using that the adjoints on $\cal H$ of $(h_s)_{AB},\hat{V}$ are 
given by $\overline{(h_s)_{AB}},\hat{V}$ respectively and the unitarity 
of $SU(2)$ we compute 
\ba \label{24}
(\hat{O}_s^{AB})^\dagger&=&[\hat{V},\overline{(h_s^{-1})_{CB}}]
\overline{(h_s)_{AC}} \nonumber\\
&=& -[(h_s)_{BC},\hat{V}](h_s^{-1})_{CA} \nonumber\\
&=&-([(h_s)_{BC}(h_s^{-1})_{CA},\hat{V}]-(h_s)_{BC}[(h_s^{-1})_{CA},\hat{V}])
\nonumber\\
&=& +\hat{O}_s^{BA}\;.
\ea
Next, using that $\underline{X}^I_s$ commutes with $\hat{O}_s^{AB}$ and that 
$(\underline{X}^I_s)^\dagger=-\underline{X}^I_s$ is anti-self-adjoint on 
$\cal H$ we find for $\hat{O}_s^{AB,I}:=\hat{O}_s^{AB}\underline{X}^I_s$ that
$(\hat{O}_s^{AB,I})^\dagger=-\hat{O}_s^{BA,I}$. \\
Putting everything together we find that when setting $N=1$, $\hat{H}^E_{YM}
=:\sum_v k_v^E\hat{H}^E_{YM,v}$ with $k_v^E>0$ that 
\ba \label{25}
\hat{H}^E_{YM,v} &=&-\sum_{v\in s\cap s'}\mbox{tr}(\hat{O}_s\hat{O}_{s'})
\underline{X}_s^I \underline{X}_{s'}^I
\nonumber\\
&=&\sum_{v\in s\cap s'}\sum_{A,B;I}(\hat{O}_s^{AB,I})^\dagger
\hat{O}_{s'}^{AB,I}
=\sum_{A,B;I}[\sum_{v\in s}\hat{O}_s^{AB,I}]^\dagger
[\sum_{v\in s}\hat{O}_s^{AB,I}]
\ea
whence we have displayed $\hat{H}$ in the form of a sum of operators of 
the form $\hat{A}^\dagger\hat{A}$, that is, it has positive semi-definite and
symmetric projections. Since these projections map ${\cal 
H}_{\gamma,\vec{\lambda}}$ into itself (the dependence on the segments
$s(e)$ involved in the definition of $\hat{H}^E_{YM}$ drops out because 
of gauge invariance, see \cite{16}), it follows that the family of 
projections defines a positive and symmetric operator on $\cal H$ which 
therefore has self-adjoint extensions (actually, by methods similar as 
for the volume, area and length operators \cite{13,14,15,16} one can show 
that each projection is essentially self-adjoint and so is the whole 
operator, therfore the extension is unique). Notice also that the electric 
part does not depend on the regulator $\delta$ any longer so that the 
continuum 
limit is already taken. It therefore projects to a {\em strongly 
diffeomorphism invariant} 
self-adjoint and positive semi-definite operator on 
the Hilbert space defined by (\ref{23b}) by general theorems proved in
\cite{4}.\\
We mention that it is conceivable that the positivity of the electric piece
could be sufficient to establish positivity of the full Yang-Mills 
Hamiltonian if the Kato condition applies to the symmetrically ordered
(but not positive) magnetic piece.

\subsection{Anomaly-Freeness}

We will understand the term ``Anomaly-free" in the sequel to mean that
we have a consistent quantum theory, namely, the commutator of two
Hamiltonian constraints vanishes when evaluated on a diffeomorphism
invariant distribution. We do not mean that the commutator equals
a certain operator that is proportional to a diffeomorphism generator.
The difficulty in achieving this more ambitious goal is two-fold as 
pointed out already in \cite{6} : First, not even the generator of a 
one parameter family of diffeomorphisms exists in the representation 
that we have chosen since the associated representation of the 
diffeomorphism group does not act strongly continuously on the Hilbert 
space. Secondly, since $\{H(M),H(N)\}=V_a(q^{ab}(M_{,b}N-M N_{,b})$
where $V_a$ is the classical generator of the diffeomorphism group,
one needs to make sense out of an operator that somehow corresponds to
$q^{ab}$ which is not at all obvious to construct. 
By the structure theorem proved in the previous section we know that there 
is a chance that
there exists a well-defined operator corresponding to the {\em product}
$q^{ab}V_b$. Indeed, in \cite{18} such an operator is constructed and 
part of it actually does generate diffeomorphisms ! However, the way that
it results from computing the commutator is rather non-standard and has 
to do with reasons deeply rooted in the structure of the Hilbert space 
$\cal H$. Basically, one can show that the commutator is weakly 
equivalent to that operator because both are zero operators on 
diffeomorphism invariant states. 

This is almost equivalent to showing anomaly-freeness in the weak sense 
as stated above with which we content ourselves here and 
which by itself is also a non-trivial task. The computations are very
similar to the vacuum case so that we refrain from displaying all the 
details. The interested reader is referred to \cite{6} to fill the gaps.\\
To begin with, notice that when evaluated on a cylindrical function, the 
Hamiltonian constraint of both gravity and matter is a sum of terms of the 
structure $N_v \hat{H}_v$ where $\hat{H}_v$ is an operator built from\\ 
1) holonomies of segments of the underlying graph which start at the vertex
$v$\\
2) gravitational Volume operators which act only on holonomies along 
segments starting at $v$\\
3) Yang-Mills Laplacian operators which act only on 
holonomies along  segments starting at $v$\\
4) Fermion field and Higgs field derivatives which act only on fields 
located at $v$\\
5) Fermion field and Higgs field insertions at the vertices of $\gamma$.\\
The crucial point is that all the terms involved in $\hat{H}_v$ involve a 
factor of the form $\hat{V}_v$ or more generally $\hat{V}_v^n,\;n>0$ where 
again this notation means the volume 
operator for an arbitrarily small neighbourhood of the vertex $v$. 
Notice that if $f$ is a function cylindrical with respect to a graph $\gamma$
then $\hat{H}(N)f=\hat{H}_\gamma(N)f=\sum_{v\in V(\gamma)} 
N_v\hat{H}_{v,\gamma} f$ 
is in general a function cylindrical with respect to a
graph $\gamma(v)$ which contains additional vertices, but these vertices 
are co-planar (arising from loop-insertions due to the gravitational or 
magnetic part of the Yang-Mills Hamiltonian constraint) or co-linear 
(arising from fermion field insertions due to the Dirac Hamiltonain 
coinstraint). Therefore, $\hat{V}_{v'}=0$ for $v'\in\gamma(v)-\gamma\forall
v\in V(\gamma)$. We therefore conclude that just like in the source free case
\ba \label{30}
&&[\hat{H}(M),\hat{H}(N)]f \nonumber\\
&=& \sum_{v,v'\in V(\gamma)}(M_{v'} N_v-M_v 
N_{v'}) \hat{H}_{\gamma(v),v'}\hat{H}_{\gamma,v}f\nonumber\\
&=& \frac{1}{2}\sum_{v,v'\in V(\gamma)}(M_{v'} N_v-M_v N_{v'})
(\hat{H}_{\gamma(v),v'}\hat{H}_{\gamma,v}
-\hat{H}_{\gamma(v'),v}\hat{H}_{\gamma,v'})f\nonumber\\
&=& \frac{1}{2}\sum_{v\not=v',v,v'\in V(\gamma)}
(M_{v'} N_v-M_v N_{v'})
(\hat{H}_{\gamma(v),v'}\hat{H}_{\gamma,v}
-\hat{H}_{\gamma,v'}\hat{H}_{\gamma(v'),v})f
\nonumber\\
\ea
In the last step we used that 
the terms with $v=v'$ trivially vanish while for the terms with $v\not=v'$
the local character of the operator $\hat{H}_{\gamma,v}$, performing 
changes of $\gamma$ only in a neighbourhood of $v$ makes it commute with 
$\hat{H}_{\gamma(v),v'}$. It is then easy to see, by the same argument as in
\cite{6}, that the two functions 
$\hat{H}_{\gamma(v),v'}\hat{H}_{\gamma,v}f,\;
\hat{H}_{\gamma,v'}\hat{H}_{\gamma(v'),v}f$ are related by an analyticity 
preserving diffeomorphism.\\
This furnishes the proof of anomaly-freeness.

\subsection{Solutions of the Diffeomorphism and Hamiltonian constraint}

The general solution of the Diffeomorphism constraint 
for theories including Fermions and Higgs fields is constructed in 
\cite{TTKin}. As for the pure gauge field case \cite{4}, they are 
elements of $\Phi'$.

We can now look for solutions to the Hamiltonian constraint. Recall 
(\cite{6}) that a solution to
the Hamiltonian and diffeomorphism constraint is a distribution 
$\Psi\in\Phi'$, that is, a continuous linear functional on the space $\Phi$,
the finite linear combinations of spin-colour-network states such that\\
1) 
$\Psi[\hat{U}(\varphi)f]=\Psi[f]\;\forall\;\varphi\in\mbox{Diff}(\Sigma),\;
f\in\Phi$ and \\
2)$\Psi[\hat{H}(N)f]=0\;\forall\;N\in{\cal S}(\Sigma),\;
f\in\Phi$.\\
Here, as usual, $\cal S$ is the Schwartz space of functions of rapid 
decrease.\\
The complete set of solutions can now be precisely characterized as 
follows :\\
Let $R:=\cup_{N\in{\cal S}}\mbox{Ran}(\hat{H}(N))$ be the union of 
the ranges of $\hat{H}(N)$ on $\Phi$ and let $S:=R^\perp$ be its 
orthogonal complement in $\Phi$. Next, for each $s\in S$ there is a 
decomposition
$s=\sum_I s_I T_I$ of $s$ into spin-colour-network states. Consider for
each $T_I$ its orbit 
$\{T_I\}:=\{\hat{U}(\varphi)T_I;\;\varphi\in\mbox{Diff}(\Sigma)\}$ under 
diffeomorphisms and construct the distribution 
$[T_I]:=\sum_{T\in\{T_I\}} T\in \Phi'$. That this is still an element of 
$\Phi'$ follows from the fact that above we took the orthogonal complement
in $\Phi$ and not in $\cal H$. Define $[s]:=\sum_I s_I [T_I]$.
Then the the complete space of solutions to both constraints is given by
${\cal V}_{phys}$ the (infinite) linear combinations of elements of the 
set  $\{[s]\; s\in S\}\subset\Phi'$.\\
We see that we know the space of solutions once we know $S$. A compact 
algorithm to describe $S$ as for the source-free case is not 
available at the moment for the matter coupled case so we restrict 
ourselves to displaying
trival solutions which is enough to show that the theory is not empty :\\
Trivial solutions are, for instance, distributions which are defined by 
graphs all of whose vertices are co-planar, just because the volume operator 
annihilates such states so that they cannot be in the image of the 
Hamiltonian constraint. The dependence on Fermion and Higgs Fields of these
solutions is completely arbitrary.\\
While we do not have an explicit algorithm for the construction of the 
kernel at the moment, it is clear that
it exists and therefore one can construct a physical inner product and 
strict Dirac observables as outlined in \cite{18}.\\
\\
\\ 
\\ 
{\large Acknowledgements}\\
\\
This research project was supported in part by DOE-Grant
DE-FG02-94ER25228 to Harvard University.

\begin{appendix}

\section{Canonical real-connection formulation of Einstein-Dirac Theory}

In this section we wish to derive the canonical action principle for 
general relativity coupled to Dirac fields in manifestly real form and in 
terms of the canonical pair $(A_a^i=\Gamma_a^i+K_a^i,E^a_i)$ introduced 
in section 2. This has not been done so far in the literature because either
one was interested in the standard Palatini formulation or in the Ashtekar
formulation, the latter involving complex-valued connections which are
difficult to deal with in the quantization programme. The progress that 
one is able 
to make with the real-valued variables for the source-free case \cite{6}
motivate to derive a similar form of the action in the matter-coupled case.

Let us begin with the (massfree) Dirac action in covariant form 
\be \label{a1}
S_{Dirac}=\frac{i}{2} \int_M d^4x 
\sqrt{-\det(g)}[\overline{\Psi}\gamma^\alpha
\epsilon^a_\alpha\nabla_a\Psi-\overline{\nabla_a\Psi}\gamma^{\alpha}
\epsilon^a_\alpha\Psi]
\ee
where $\gamma^\alpha$ are the Minkowski space Dirac matrices, 
$\epsilon^a_\alpha$ are the tetrad fields and $\Psi=(\psi,\eta)$ is a Dirac
bi-spinor and $\overline{\Psi}=(\Psi^\star)^T\gamma^0$ its conjugate. Here 
$\psi=(\psi^A)$ and $\eta=(\eta_{A'})$ 
transform according to the fundamental representations of $SL(2,\Co)$ and 
are scalars of density weight zero.
The covariant derivative $\nabla_a$ is defined to annihilate the tetrad 
$\epsilon^a_\alpha$, that is, we are using the second order formalism.\\
In order to put (\ref{a1}) into canonical form we take $M=\Rl\times\Sigma$,
let $T^a$ be the time foliation vector field of $M$ and denote by $n^a$ the 
normal vector field of the time slices $\Sigma$. Then the tetrad can be 
written $\epsilon^a_\alpha=e^a_\alpha-n^a n_\alpha$ with 
$e^a_\alpha n_a=e^a_\alpha n^\alpha=0$ so that $e^a_\alpha$ is a triad 
and $\eta^{\alpha\beta}n_\alpha\n_\beta=-1$ is an internal unit timelike 
vector which we may choose to be $n_\alpha=-\delta_{\alpha,0}$
($\eta=\mbox{diag}(-,+,+,+)$ is the Minkowski metric). Finally,
inserting lapse and shift fields by $(\partial_t)^a=T^a=N n^a+N^a$ with $N^a 
n_a=0$ one 
sees that the action can be written, after lengthy computations, in terms of 
Weyl spinors as (using 
the Weyl representation for the Dirac matrices, for instance, to expand out
various terms) 
\ba \label{a2}
S_{Dirac}&=&\frac{i}{2}\int_\Rl dt\int_\Sigma d^3x N\sqrt{\det(q)}
[\frac{T^a-N^a}{N}(\psi^\dagger\nabla_a^+\psi+\eta^\dagger\nabla_a^-\eta-c.c.)
\nonumber\\
&&+e^a_i(\psi^\dagger\sigma_i\nabla_a^+\psi-\eta^\dagger\sigma_i\nabla_a^
-\eta-c.c.)]
\ea
where the $c.c.$ in $(*+c.c.)$ stands for ``complex conjugate of $*$.  
Here we have defined $e^a_\alpha=(0,e^a_i)$ and abused the notation in 
writing $e^a_i=(e^t_i=0,\mbox{sgn}(\det(e))e^a_i)$,\footnote{the sign
factor is a possible choice because it does not appear in $q_{ab}=e_a^i
e_b^i$} 
$\sigma_i$ are the Pauli matrices,
$\psi^\dagger:=(\psi^\star)^T$ and $\nabla_A^\pm$ is the self-dual 
respectively
anti-self-dual part of $\nabla_a$ in the Weyl representation. More precisely,
$\nabla^\pm_a=\partial_a+\omega_a^\pm,\;\omega^\pm_a=-i\sigma_j
\omega_a^{j\pm},\; \omega_a^{j+}=-1/2\epsilon_{jkl}\omega_a^{kl+},\;
\omega_a^{\alpha\beta+}=\frac{1}{2}(\omega_a^{\alpha\beta}
-i\epsilon^{\alpha\beta}\;_{\gamma\delta} \omega_a^{\gamma\delta}),
\;\omega_a^{j-}=\overline{\omega_a^{j+}}$ and $\omega_a^{\alpha\beta}$ is 
the spin-connection of $\epsilon^a_\alpha$. The unfamiliar reader is 
referrred 
to the standard literature on the subject (for instance \cite{Wald}).\\
It is easy to see that the spatial part of $\omega_a^{j+}$
is just given by $\frac{1}{2}A_a^{j\Co}$ where $A_a^{j\Co}=\Gamma_a^j+i 
K_a^j$ is the complex-valued Ashtekar connection. Denoting 
${\cal D}_a^\Co\psi=(\partial_a+A_a^{j\Co}\tau_j)\psi$ and 
$\overline{{\cal D}}_a^\Co\eta=(\partial_a+\overline{A_a}^{j\Co}\tau_j)\eta$ 
with $\tau_j=-\frac{i}{2}\sigma_j$ (Pauli matrices) and 
$A_t^{j\Co}=T^a\omega_a^{j+},\;\dot{\psi}=
T^a\partial_a\psi$ we end up with 
\ba \label{a3}
S_{Dirac}&=&\frac{i}{2}\int dt\int d^3x\sqrt{\det(q)}
[(\psi^\dagger\dot{\psi}+\eta^\dagger\dot{\eta}-c.c.)
\nonumber\\
&& 
-(-(A_t^{j\Co}\psi^\dagger\tau_j\psi
+\overline{A_t^{j\Co}}\eta^\dagger\tau_j\eta-c.c.)
+N^a(\psi^\dagger{\cal D}_a^\Co \psi+\eta^\dagger\overline{{\cal D}}_a^\Co 
\eta-c.c.) \nonumber\\
&& 
+N \mbox{sgn}((\det(e)) e^a_i(-\psi^\dagger\sigma_i{\cal D}_a^\Co 
\psi+\eta^\dagger\sigma_i \overline{{\cal D}}_a^\Co \eta-c.c.))]\;.
\ea
Let us now introduce $D_a\psi=(\partial_a+\tau_j\Gamma_a^j)\psi,\; E^a_i
=\det(e_a^i)e^a_i,\; A_t^j:=\Re(A_t^{j\Co})$ then we see by explicitly
evaluating $c.c.$ that 
\ba \label{a4}
S_{Dirac}&=&\frac{i}{2}\int dt\int d^3x\sqrt{\det(q)}
[(\psi^\dagger\dot{\psi}+\eta^\dagger\dot{\eta}-c.c.)
\nonumber\\
&& 
-(-2A_t^j(\psi^\dagger\tau_j\psi+\eta^\dagger\tau_j\eta)
+N^a(\psi^\dagger D_a\psi+\eta^\dagger D_a\eta-c.c.) \nonumber\\
&& 
+N \frac{E^a_i}{\sqrt{\det(q)}}([-\psi^\dagger\sigma_i 
D_a\psi+\eta^\dagger\sigma_i D_a\eta-c.c.]+2[K_a,E^a]^j
(\psi^\dagger\tau_j\psi-\eta^\dagger\tau_j\eta)))]\;. \ea
This is the 3+1 split Dirac action that we are going to combine with 
the 3+1 split Einstein action to obtain the desired form in terms of 
$(A_a^i,E^a_i)$.
 
We come to the Einstein action. In contrast to \cite{Jacobson} we also take
the second order form of the Palatini action (that is, we let the 
gravitational connection be the one that annihilates the tetrad from the 
outset). Otherwise we can take over the results from \cite{Jacobson} and 
arrive at $S_{Einstein}=\Re(S_E^+)$ where $S_E^+$ is the self-dual part of
$S_{Einstein}$ which in our notation is written as 
\be \label{a5}
S_E^+=\frac{1}{\kappa}=\int dt\int d^3x[-i\dot{A}_a^{j\Co}E^a_j
-(iA_t^{j\Co}{\cal D}_a^\Co E^a_j-iN^a\mbox{tr}(F_{ab}^\Co E^b)
+\frac{N}{2\sqrt{\det(q)}}\mbox{tr}(F_{ab}^\Co[E^a,E^b])]
\ee
where $F^\Co$ denotes the curvature of $A^\Co$ and $\kappa$ the 
gravitational coupling constant. Computing the real part reveals
\ba \label{a6}
S_{Einstein}&=&\frac{1}{\kappa}=\int dt\int d^3x[\dot{K}_a^j E^a_j
-(-A_t^j[K_a,E^a]_j+2N^a D_{[a} K_{b]}^j E^b_j\nonumber\\
&&-\frac{N}{2\sqrt{\det(q)}}\mbox{tr}(([K_a,K_b]-R_{ab})[E^a,E^b])]
\ea
where $R_{ab}$ is the curvature of $e_a^i$.\\
Thus, putting both actions together, we find that the gravitational 
Gauss constraint is given by (no other matter contributes to it)
\be \label{a7}
{\cal G}_j=\frac{1}{\kappa}[K_a,E^a]_j+i\sqrt{\det(q)}[\psi^\dagger\tau_j\psi
+\eta^\dagger\tau_j\eta]\;.
\ee
We can now perform a canonical point transformation on the gravitational 
phase space given by $(K_a^i,E^a_i)\to(A_a^i,E^a_i)$ (the generator turns 
out to be $\int d^3x \Gamma_a^i E^a_i$ as one can explicitly check) and 
we must then express the constraints in terms of $A_a^i$. Let us 
therefore introduce the real-valued derivative ${\cal D}_a\psi:=
(\partial_a+A_a^j\tau_j)\psi$ and denote by $F_{ab}$ the curvature of 
$A_a^i$. Using that $D_a E^a_i=0$ we can immediately write
\be \label{a8}
{\cal G}_j=\frac{1}{\kappa}{\cal D}_a 
E^a_j+i\sqrt{\det(q)}[\psi^\dagger\tau_j\psi +\eta^\dagger\tau_j\eta]\;. 
\ee
Next, we expand $F_{ab}$ in terms of $\Gamma_a,K_a$, use the Bianchi identity
$\mbox{tr}(R_{ab}E^b)=0$ and find that the vector constraint $V_a$,
the coefficient of $N^a$ in $S_{Dirac}+S_{Einstein}$ is given, up to a term
proportional to ${\cal G}_j$, by
\be \label{a9}
V_a=\mbox{tr}(F_{ab} E^b)+\frac{i}{2}\sqrt{\det(q)}(\psi^\dagger
{\cal D}_a\psi +\eta^\dagger{\cal D}_a\eta-c.c).
\ee
Finally, let as in the source-free case
\be \label{a10}
H_E=\frac{1}{2\kappa}\mbox{tr}(F_{ab}\frac{[E^a,E^b]}{\sqrt{\det(q)}})
\ee
which has the interpretation of the source-free Euclidean 
Hamiltonian constraint. Furthermore, let 
\be \label{a11}
H_G:=-H_E+\frac{2}{2\kappa}
\mbox{tr}([K_a,K_b]\frac{[E^a,E^b]}{\sqrt{\det(q)}})
\ee
which in the source-free case would be the full Lorentzian Hamiltonian 
constraint. Then the Einstein contribution to Hamiltonian constraint
of $S_{Dirac}+S_{Einstein}$ is given by
\be \label{a12}
H=H_G-\frac{2}{2\kappa}D_a\mbox{tr}([K_b,E^b]\frac{E^a}{\sqrt{\det(q)}})
=:H_G+T\;.
\ee
Notice that in the source-free case the correction $T$ of $H$ to $H_G$ is 
proportional to a Gauss constraint and therefore would vanish separately 
on the constraint surface. However, in our case, using the Gauss 
constraint (\ref{a7}) we find that
\be \label{a13} 
T=-\frac{1}{2}([K_a,E^a]^j-E^a_j{\cal D}_a)J_j
\ee
where we have defined the current $J_j=\psi^\dagger\sigma_j\psi+\eta^\dagger
\sigma_j\eta$. On the other hand, writing also the Dirac contribution to
the Hamiltonain constraint in terms of ${\cal D}_a$ rather than $D_a$ 
and combining with $H$ we find that the first term on the right hand side 
of (\ref{a13}) cancels against a similar term. We end up with the 
contribution $C$ from both the Einstein and Dirac sector to the 
Hamiltonian constraint which is given, up to a term proportional to the 
gravitational Gauss constraint, by
\ba \label{a14}
C&=&H_G+\frac{E^a_j}{2\sqrt{\det(q)}}({\cal D}_a(\sqrt{\det(q)})J_j)
\nonumber\\
&&+i\sqrt{\det(q)}
[\psi^\dagger\sigma_j{\cal D}_a\psi-\eta^\dagger\sigma_j{\cal D}_a\eta-c.c.]
-K_a^j\sqrt{\det(q)}(\psi^\dagger\psi-\eta^\dagger\eta))\;.
\ea
In order to arrive at (\ref{a14}) one has to use the Pauli matrix algebra
$\sigma_j\sigma_k=\delta_{jk}1_{SU(2)}+i\epsilon_{jkl}\sigma_l$ at several 
stages when computing $c.c.$ Notice that we can write (\ref{a14}) also in 
terms of the half-densities $\xi=\root 4\of{\det(q)}\psi,\;\rho=
\root 4\of{\det(q)}\eta$ by absorbing the $\sqrt{\det(q)}$ appropriately
and using that ${\cal D}_a\det(q)=0$. We find
\ba \label{a15}
C&=&H_G+\frac{E^a_j}{2\sqrt{\det(q)}}({\cal D}_a(
\xi^\dagger\sigma_j\xi+\rho^\dagger\sigma_j\rho)
\nonumber\\
&&+i
[\xi^\dagger\sigma_j{\cal D}_a\xi-\rho^\dagger\sigma_j{\cal 
D}_a\rho-c.c.] -K_a^j(\xi^\dagger\xi-\rho^\dagger\rho))\;.
\ea
Note also that 
$i\sqrt{\det(q)}[\psi^\dagger\dot{\psi}-\dot{\psi}^\dagger\psi]
=i[\xi^\dagger\dot{\xi}-\dot{\xi}^\dagger\xi]$ so that our change of 
variables is actually a symplectomorphism !\\
This is the form of the constraint that we have been looking for : 
up to $K_a^i$ we have expressed everything in terms of real-valued 
quantities and the canonically conjugate pairs 
$(\xi,i\overline{\xi}),(\rho,i\overline{\rho})$. Now, let as in the 
source-free case denote $V=\int d^3x \sqrt{\det(q)}$ the total volume of 
$\Sigma$ and $H_E(1)=\int d^3x H_E(x)$. Then it is still true that 
$K_a^j=-\{A_a^j,\{V,H_E(1)\}\}$ and since $V,H_E(1)$ admit well-defined 
quantizations \cite{6} we conclude that despite its complicated appearance
(\ref{a15}) admits a well-defined quantization as well.
Note that if we had not worked with half-densities $\xi,\rho$ but with the 
$\psi,\eta$ then, while $i\sqrt{\det(q)}\bar{\psi}$ is the momentum conjugate
to $\psi$, the gravitational connection would get a correction 
proportional to $i e_a^i [\psi^\dagger\psi+\eta^\dagger\eta]$. Thus we
would have had to admit a complex connection which would be 
desasterous as the Hilbert space techniques \cite{0,1,2,3,3a,4} would not 
be at our disposal. Therefore the strategy of working with $\psi$'s
as advertized in \cite{21} is not suitable for quantizing the 
Einstein-Dirac theory and we are forced to adapt the framework developed in
\cite{TTKin}.

\end{appendix}

\end{document}